\documentclass{aa}  

\usepackage{graphicx}
\usepackage{txfonts}
\usepackage{xcolor}
\newcommand{\veloce}{{\texttt{VELOCE}}}

\begin{document}

   \title{The VELOCE modulation zoo II.}
   \subtitle{Humps and splitting patterns in spectral lines of classical Cepheids}

\author{H. Netzel\thanks{henia@netzel.pl}
          \and
          R. I. Anderson
          \and
          G. Viviani
          }

   \institute{Institute of Physics, \'Ecole Polytechnique F\'ed\'erale de Lausanne (EPFL), Observatoire de Sauverny, 1290 Versoix, Switzerland}

   \date{Received \today; accepted ...}


\abstract
{Line splitting in spectral lines is observed in various types of stars due to phenomena such as shocks, spectroscopic binaries, magnetic fields, spots, and non-radial modes. In pulsating stars, line splitting is often attributed to pulsation-induced shocks. However, this is rarely observed in classical Cepheids, with only a few reports, including X Sagittarii and BG Crucis, where it has been linked to atmospheric shocks.}
{We investigate line splitting in X Sgr and BG Cru using spectroscopic time series, and search for similar phenomena in other classical Cepheids.} 
{High signal-to-noise cross-correlation function (CCF) time series from the VELOcities of CEpheids (\veloce) project are analyzed. This dataset spans several years, allowing us to study the periodicities and evolution of CCF features. For X Sgr and BG Cru, we perform a detailed analysis of the individual components of the split CCFs. Additionally, we search for periodicities in CCF variations and examine other classical Cepheids for distortions resembling unresolved line splitting.} 
{We confirm line splitting in X Sgr and BG Cru, trace the features over time, and uncover the periodicity behind them. Several other Cepheids also exhibit CCF humps, suggesting unresolved or marginally resolved line splitting. We discuss the incidence and characteristics of these stars.}
{The periodicity of line splitting in X Sgr and BG Cru differs significantly from the dominant pulsation period, ruling out pulsation-induced shocks. The periodicities are too short for rotation-related phenomena, suggesting non-radial modes as the most likely explanation, though their exact nature remains unknown. We also identify humps in six additional stars, indicating an incidence rate of 3\% in the \veloce\ sample.}

   \keywords{Stars: oscillations (including pulsations) --
                Stars: variables: Cepheids --
                Techniques: radial velocities
               }

   \maketitle
%

\section{Introduction}

Classical Cepheids are intermediate-mass pulsating stars that evolve through the classical instability strip during core He burning phase. Typically they pulsate in one or two low order radial modes. Classical Cepheids pulsating in three radial modes simultaneously are also known however such stars are not as common \citep{soszynski2015}. The importance of Cepheids stems from the fact that they obey a period-luminosity relation first discovered by \cite{leavitt1908, leavitt1912}. They serve as an important rung on the cosmic distance ladder and are used to determine the Hubble constant \citep[e.g.][]{2021ApJ...908L...6R}. Cepheids are also interesting as the evolved counterparts of intermediate-mass stars and thus can inform stellar evolution in general. However, the physics of pulsation of Cepheids is still not completely understood. For instance, such puzzling aspects are mass-luminosity relation, relating to chemical composition, opacities \citep{moskalik.buchler1992}, convective core overshooting \citep[e.g.][]{bono2000, marconi2005}, pulsation-enhanced mass-loss \citep{neilson.lester2008}, and rotation \citep{anderson2014}, among other things. New and stronger constraints on stellar structure are urgently needed to make progress given this wide array of physical effects. Classical Cepheids show additional phenomena besides pulsations in radial modes, such as periodic or quasi-periodic modulations of pulsations \citep{smolec2017_cep, suveges.anderson2018a, suveges.anderson2018b}, or pulsations in the additional non-radial modes \citep[for a review see][and references therein]{netzel_review}. Studying such phenomena and the physics behind them could eventually lead to better understanding of classical Cepheids.

Spectroscopic time series allows to study line profile variations due to pulsations. In the case of classical Cepheids, the dominant effect are high-amplitude radial pulsations that affect the shapes and position of spectral lines due to atmospheric motions and related changes in temperature. In principle however, other phenomena, such as low-amplitude non-radial modes, shock waves, stellar spots can also manifest in line profile variations. 

In particular, shock waves passing through an atmosphere can cause splitting of spectral lines via the \citet{1952SchwarzschildMechanism} mechanism. Such features are commonly observed in other types of pulsating stars, including RR Lyrae stars \citep{fokin.gillet1997} and type II Cepheids (see e.g. \citealt{fokin.gillet1994}, and fig. 1 in \citealt{veloce} for W Vir), or Miras \citep{alvarez2001}. In these types of pulsating stars, the line splitting phenomenon was attributed to pulsation-induced shock-waves and shares a characteristic that line splitting appears at specific pulsation phases, and in synchronization with the dominant pulsation period.

Interestingly, in the case of classical Cepheids line splitting was rarely detected. \citet{kraft1956} reported line splitting in spectra of X Cyg. Line splitting in X Sgr was reported for the first time based on infrared spectra by \citet{sasselov1990}. \citet{kovtyukh2003} confirmed this observation for X Sgr, and reported similar features in three more classical Cepheids: BG Cru, EV Sct, and V1334 Cyg. The above mentioned detections of line splitting in several classical Cepheids were reported based on individual spectral lines and few observations. In principle, however, cross-corellation function (CCF) profiles, which represent a weighted average line profile, should also show these features if they are present in individual lines. Indeed, \citet{anderson2013} reported line splitting in CCF profiles of three classical Cepheids X Sgr, BG Cru, and LR TrA. 

The origin of line splitting in X Sgr is uncertain. \cite{kovtyukh2003} proposed that the observed features in line profile variations are due to additional pulsations in non-radial modes. On the other hand, \cite{mathias2006} studied in detail X Sgr using high-resolution spectroscopy and preferred an explanation involving multiple shock-waves for the observed line splitting. Conversely, \cite{anderson2013} collected several individual CCF profiles at the same pulsation phases of consecutive pulsation cycles for X Sgr and showed that the CCFs appear significantly different. This is not expected in the context of the pulsation-induced shock wave phenomenon as a cause for line splitting. Other phenomenon that may distort line profiles in similar way, besides shocks and non-radial modes are spots which cause humps traveling across the line profiles due to stellar rotation \citep[see e.g. fig. 2.3 in][]{semenova2006}. In this scenario, the variations of the traveling hump is periodic and related to the rotation period. Whether this might be the case for X Sgr was not yet explored.

Here we use spectroscopic observations collected as part of the VELOcities of CEpheids project \citep[\veloce,][]{veloce} to investigate in detail the line splitting observed in X Sgr and BG Cru. Moreover, we search for more stars showing similar features. 

In Sec.~\ref{sec:methods} we describe used data and methodology. Results are presented in Sec.~\ref{sec:results} and discussed in Sec.~\ref{sec:discussion}. Sec.~\ref{sec:conculsion} contains conclusions.

\section{Data and methods}\label{sec:methods}

The first data release of the \veloce\ project has published over 18,000 precise radial velocities of 258 galactic Cepheids from both hemispheres \citep{veloce}. The spectroscopic observations were carried out with the Hermes high-resolution ($R \sim 85\,000$) spectrograph  \citep{raskin2011} mounted on the Flemish Mercator Telescope at Roque de los Muchachos Observatory (Spain) for the northern targets, and with the Coralie high-resolution ($R \sim 60\,000$) spectrograph \citep{queloz2001_coralie} mounted on the 1.2m Swiss Euler Telescope at La Silla (Chile) for the southern targets. 

Both instruments have undergone upgrades that alter the optical path and hence the instrumental line shapes. We split the datasets according to the times of these interventions into the Coralie07 and Coralie14 data set for the Coralie instrument, following \cite{veloce}. Hermes underwent upgrades on April 25th 2018. We refer to the data collected after this intervention as Hermes18.

\veloce\ employs cross-correlation functions (CCF) to increase precision of radial velocity (RV) measurements  \citep{baranne1996,pepe2002}. The correlation mask chosen for the CCF computations represents an approximately solar-metallicity solar-like (G2) star and contains a few thousand metallic absorption lines. Such approach significantly increases the signal-to-noise of CCFs compared to individual lines at the cost of reducing the information provided from individual lines and of introducing weighting according to the correlation mask.

We used CCF profiles to trace the line splitting and performed a frequency analysis of the time-series of RV, and CCFs' shape indicators. We used the full-width at half-maximum (FWHM), equivalent width (EW), bisector inverse span (BIS), and relative depth (contrast). These parameters describe well the whole shape of the CCF. In particular, FWHM measures the broadening of the profile. EW and contrast measure its strength. BIS is defined as the velocity difference between the midpoints of horizontal lines placed near the top and bottom of the profile. It effectively measures the asymmetry of the profile. It also benefits from the fact that, as a velocity difference, it is not affected by wavelength scale uncertainty as in the case of RV. BIS was already successfully used to detect additional long-period signals in Polaris \citep{anderson2019_polaris}. RV with the shape parameters provide detailed view on CCF changes due to pulsations and other unexpected phenomena \citep[e.g.,][]{anderson2016_masks}. Analysis of shape indicators was already proven successful in detection of signals in classical Cepheids that are likely due to non-radial modes \citep{netzel_veloce}. In this study we analyze these parameters for the two main targets with the most prominent line splitting: X Sgr and BG Cru, and for the other candidates that might show line-splitting (see Sec.~\ref{Subsec:candidates} for further details).

X Sgr was observed with both instruments, Coralie and Hermes, while BG Cru was observable only with Coralie. The dataset is summarized in Table~\ref{tab:data_summary}.  The data for BG Cru for Coralie14 was published by \cite{netzel_veloce}. In Table~\ref{tab:data_sample} we publish Coralie07 data for BG Cru, Coralie07, Coralie14, and Hermes data for X Sgr, and Hermes and Hermes18 data for SZ Cas. Note some missing values that were removed during the analysis as significant outliers. We note that the dataset published by \cite{veloce} includes the observations until 5th of March 2022, i.e. BJD=2459644. Since observations for \veloce\ are ongoing, we included in the present analysis the most recent observations as well.

\begin{table}[]
    \caption{Summary of datasets used for the analyzed targets.}
    \centering	
    \begin{tabular}{lllll}
       ID & BJD range & <S/N> & N & Instr. \\ 
       \hline
       X Sgr  &  2456090--2457575 & 292 & 81 & Hermes  \\
       X Sgr  &  2456334--2456792 & 506 & 86 & Coralie07  \\
       X Sgr  &  2457152--2459820 & 1077 & 157 & Coralie14  \\
       BG Cru & 2455944--2456670 & 1557 & 50 & Coralie07 \\
       BG Cru & 2457154--2460157 & 896 & 405 & Coralie14 \\
       SZ Cas & 2455894--2457700 & 104 & 38 & Hermes \\
       SZ Cas & 2459794--2460341 & 309 & 96 & Hermes18 \\
       &
    \end{tabular}
    \tablefoot{
Consecutive columns provide ID of a target, BJD range of observations, average signal-to-noise of CCFs (S/N) calculated with FAMIAS tools \protect\citep{zima2008}, number of CCFs, and instrument of observations.
}

    \label{tab:data_summary}
\end{table}

\begin{table*}
    \caption{Sample of a table containing the analyzed data for X Sgr, BG Cru, and SZ Cas.}
    \centering
    \begin{tabular}{llllllll}
    ID & BJD-2.4M & RV [km\,s$^{-1}$] & RV$_{\rm err}$[km\,s$^{-1}$] & FWHM [km\,s$^{-1}$] & BIS [km\,s$^{-1}$] & Contrast [\%] & EW [km\,s$^{-1}$] \\
    \hline
X Sgr	&	56090.65896	&	0.03137	&	0.411	&	40.57355	&	--23.57676	&		&	5.15696	\\
X Sgr	&	56091.59041	&	--10.67392	&	0.522	&	44.10578	&	--1.37119	&		&	4.29281	\\
X Sgr	&	56092.59433	&	--30.35745	&	0.348	&	37.13551	&	5.15865	&		&	3.32875	\\
\vdots	&	\vdots	&	\vdots	&	\vdots	&	\vdots	&	\vdots	&	\vdots	&	\vdots	\\
SZ Cas	&	55894.68323	&	--38.38718	&	0.117	&	26.65177	&	--2.12898	&	22	&	5.77734	\\
SZ Cas	&	55894.68718	&	--38.38318	&	0.119	&	27.07223	&	--1.97319	&	21	&	5.75281	\\
SZ Cas	&	55895.64345	&	--37.522	&	0.114	&	25.29172	&	--1.53166	&	21	&	5.68855	\\
\vdots	&	\vdots	&	\vdots	&	\vdots	&	\vdots	&	\vdots	&	\vdots	&	\vdots	\\
BG Cru	&	55944.8515	&	--17.53361	&	0.006	&	29.85759	&	--1.07677	&	14.367	&	4.35791	\\
BG Cru	&	55945.76954	&	--14.27847	&	0.008	&	29.97627	&	--1.92032	&	13.476	&	4.25527	\\
BG Cru	&	55946.8011	&	--23.87647	&	0.012	&	30.0712	&	1.6422	&	12.44	&	3.86342	\\
\vdots	&	\vdots	&	\vdots	&	\vdots	&	\vdots	&	\vdots	&	\vdots	&	\vdots	\\
&
    \end{tabular}
    \tablefoot{Consecutive columns provide star’s name, BJD of observation, RV with its error, FWHM, BIS, contrast, and EW values. Full table is available at the CDS.}
    \label{tab:data_sample}
\end{table*}

\subsection{Analysis of CCF profiles}\label{sec:analysisCCF}

We investigated line splitting present in X Sgr and BG Cru using three different approaches. The first approach is a version of a methodology of \cite{mathias2006} who modeled line profiles of X Sgr by simultaneous fitting of three independent Gaussian components. The observations used by \cite{mathias2006} were carried out consecutively and covered almost 1.5 pulsation cycle. Therefore they were able to trace individual components with individual Gaussians. Our dataset spans a much larger time range, including observations from dedicated \veloce\ observing runs that provide observations over up to two full consecutive cycles while also containing many long (months to seasons) gaps. Consequently, we had very limited possibility to trace individual components.

Following \cite{mathias2006} we represent the CCFs of X Sgr using a triple Gaussian fit, and we find that BG Cru also requires three Gaussians to achieve a good fit, despite an apparently weaker line splitting pattern. The gaps in the data set made it unfortunately not possible to track each component's movements unambiguously as a function of time. We therefore performed time-series Fourier analysis of the red-shifted, central, and blue-shifted components separately.

\begin{figure}
    \centering
    \includegraphics[width=\columnwidth]{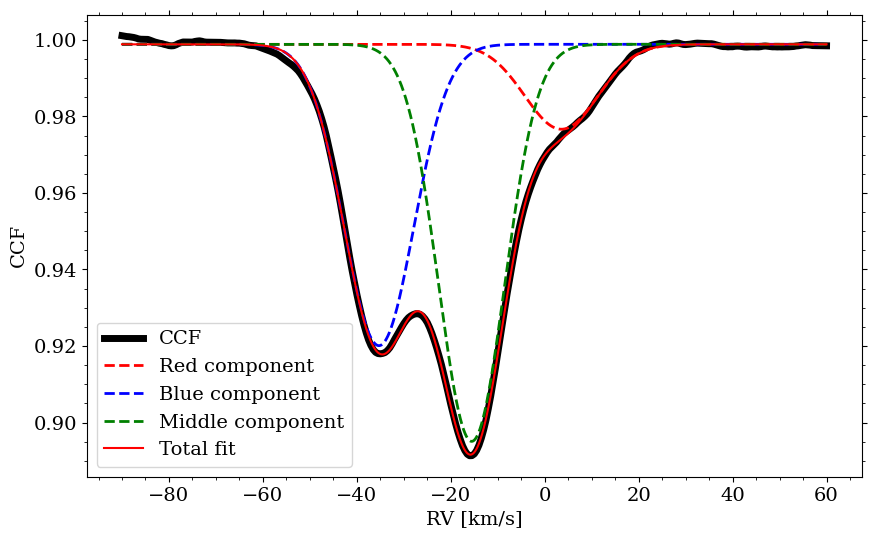}

    \includegraphics[width=\columnwidth]{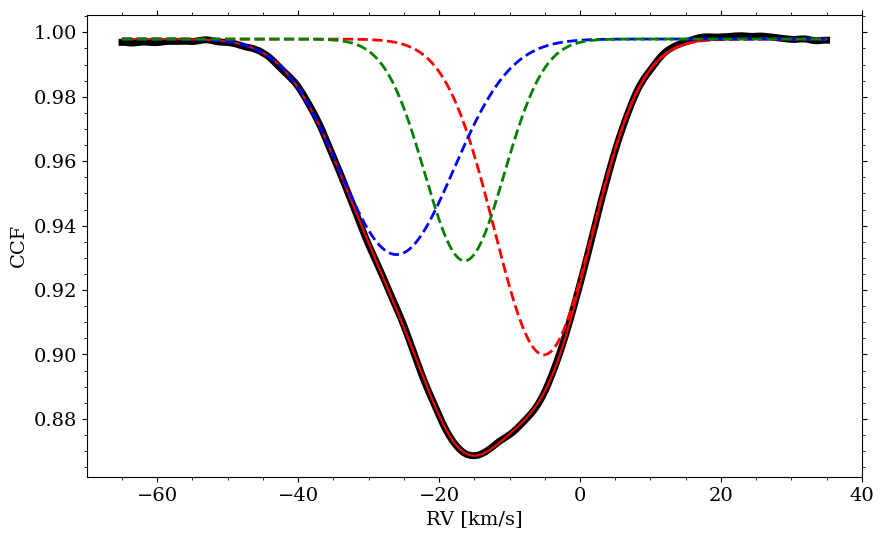}
    \caption{An example of a triple-Gaussian fit to one of the CCF profiles for X Sgr  (top panel) and BG Cru (bottom panel).}
    \label{fig:gaussian_fit}
\end{figure}

In Fig.~\ref{fig:gaussian_fit} we show an example of triple-Gaussian fit to one of the CCF profiles for X Sgr and BG Cru using our dataset. We note that sometimes, when the line splitting was not significant, not all three components were fitted to the CCF profile. We excluded such instances from further analysis, as the CCFs were not well reproduced by the fit for these instances, and we made sure that all phases of pulsation are well represented. We used the parameters (depth, centroid, width) of the three Gaussians (bluest, middle, reddest) as nine time series. These time series were then analyzed using standard frequency analysis techniques. 

\begin{figure}
    \centering
    \includegraphics[width=\columnwidth]{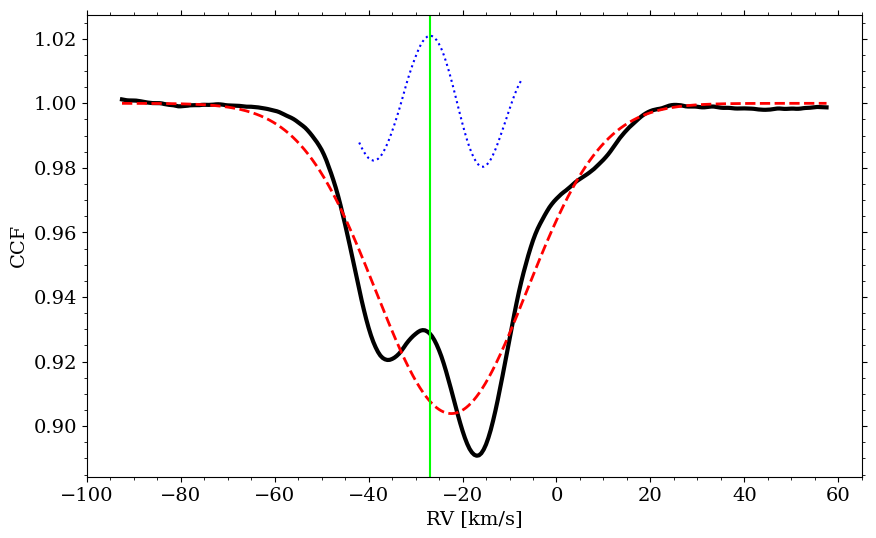}

    \includegraphics[width=\columnwidth]{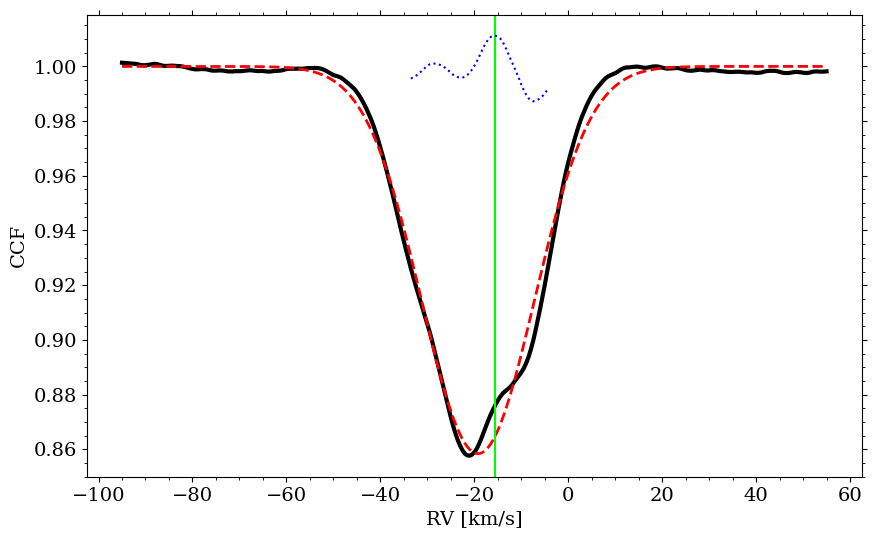}
    \caption{Example of tracing a hump on the CCF profile for X Sgr (top panel) and BG Cru (bottom panel).}
    \label{fig:pypek}
\end{figure}

In the second approach, we traced the hump formed between the two components of the line relative to the mean RV of the CCF. The position of the hump is marked with a green vertical line in Fig.~\ref{fig:pypek}. We fitted a single Gaussian profile to each CCF (red dashed line in Fig.~\ref{fig:pypek}), which provided the RV determination. The RV of the hump was identified as the maximum of the residuals (blue dotted line above the CCF profile in Fig.~\ref{fig:pypek}). The relative RV of the hump was identified as the difference between the maximum of the residuals and the RV of the CCF profile from the Gaussian fit. We then constructed a time series representing the measured relative RV of the hump and analyzed it using standard time-series analysis technique. The top panel of Fig.~\ref{fig:pypek} presents this approach for X Sgr, while the bottom panel presents it for BG Cru. Note, that in the case of X Sgr, we trace only the dominant hump that distorts the center of the CCF, even if there appear to be another hump at the wing.

In the third approach we analyzed CCFs of X Sgr and BG Cru with FAMIAS \citep{zima2008}. We studied two-dimensional frequency spectra, which allow to see how the power is distributed along the CCF profile (see Fig.~\ref{fig:2d_1d_xsgr}). We also performed consecutive prewhitening using one-dimensional average pixel-by-pixel spectra, which are calculated from the two-dimensional spectra by averaging the amplitude across the profile for each frequency (see the top panel of Fig.~\ref{fig:2d_1d_xsgr}). For this analysis we used only CCF profiles collected with Coralie14. We note that it was not possible to reach the noise level in the prewhitening procedure. After several prewhitening steps, the highest detected signal in the frequency spectrum of the residuals was the residual signal at the position of the main radial mode. In Sec.~\ref{sec:results} we report and discuss only the signals that were detected before this step.

\subsection{Frequency analysis}

We performed a frequency analysis of several different time series. We used RV, and CCF profile shape indicators: full-width at half-maximum (FWHM), bisector inverse span (BIS, see fig. 2 in \citealt{anderson2016}), equivalent width (EW), and relative depth (contrast) as collected by \veloce. We also constructed a few additional time series as described in Sec.~\ref{sec:analysisCCF} to specifically trace the periodicity of the hump in CCFs. 

To find the dominant periodicity we performed a standard Fourier analysis. In the case of RV, FWHM, BIS, EW and contrast, the dominant periodicity corresponded to the main pulsation mode. We subtracted the pulsation frequency and its harmonic from the data by fitting and subtracting a Fourier series in the form:

\begin{equation}
    \label{eq.series}
    F(t)=A_0 + \sum_k A_k \sin(2\pi f_k t + \phi_k),
\end{equation}
where $A_0$ is the mean value, $f_k$ is a dominant frequency of a radial mode and its harmonics, and $A_k$ and $\phi_k$ are their amplitudes and phases. Then we repeated the Fourier analysis of the residuals to search for any additional periodicities. We note, that since \veloce\ is a ground-based survey, there are also daily and yearly aliases in frequency spectra, which need to be taken into consideration when analyzing origin of the signals.

In the case of two additional time series constructed as described in Sec.~\ref{sec:analysisCCF}, i.e.  the RVs for each Gaussian component and the hump, we calculated frequency spectra to identify the dominant periodicity of the datasets. We note that in this case,  the dominant signal no longer corresponds to the pulsation frequency.

\subsection{Search for more stars showing the phenomenon}\label{Subsec:candidates}

\veloce\ collected observations for 258 classical Cepheids, and we searched for line splitting patterns among all of them. We visually inspected all available CCFs collected by \veloce\ for all of the monitored Cepheids and manually classified stars where the CCF profiles indicate presence of line splitting. We note, however, that this phenomenon is the strongest in the case of X Sgr and BG Cru. In the case of some stars in our sample, we noticed additional distortions and humps in the CCF profiles that suggest unresolved, or marginally resolved, line splitting patterns (see Figure~\ref{fig:ccfs_6stars} in Sec.~\ref{sec:results} for LR TrA, V0411 Lac, V1334 Cyg, SZ Cas, ASAS J174603-3528.1, and V1019 Cas). CCF shape indicators for V0411 Lac were already analyzed by \cite{netzel_veloce}. In this work, we performed a frequency analysis of CCF parameters for the remaining candidates and we report our results for one of them (SZ Cas, see Sec.~\ref{subsec:results_ccf_frequency_analysis} for details). The data used for this star is also included in Tables~\ref{tab:data_summary} and \ref{tab:data_sample}.

\section{Results}\label{sec:results}

\subsection{Occurrence of humps and splitting patterns}

\begin{figure*}
    \centering
    \includegraphics[width=\textwidth]{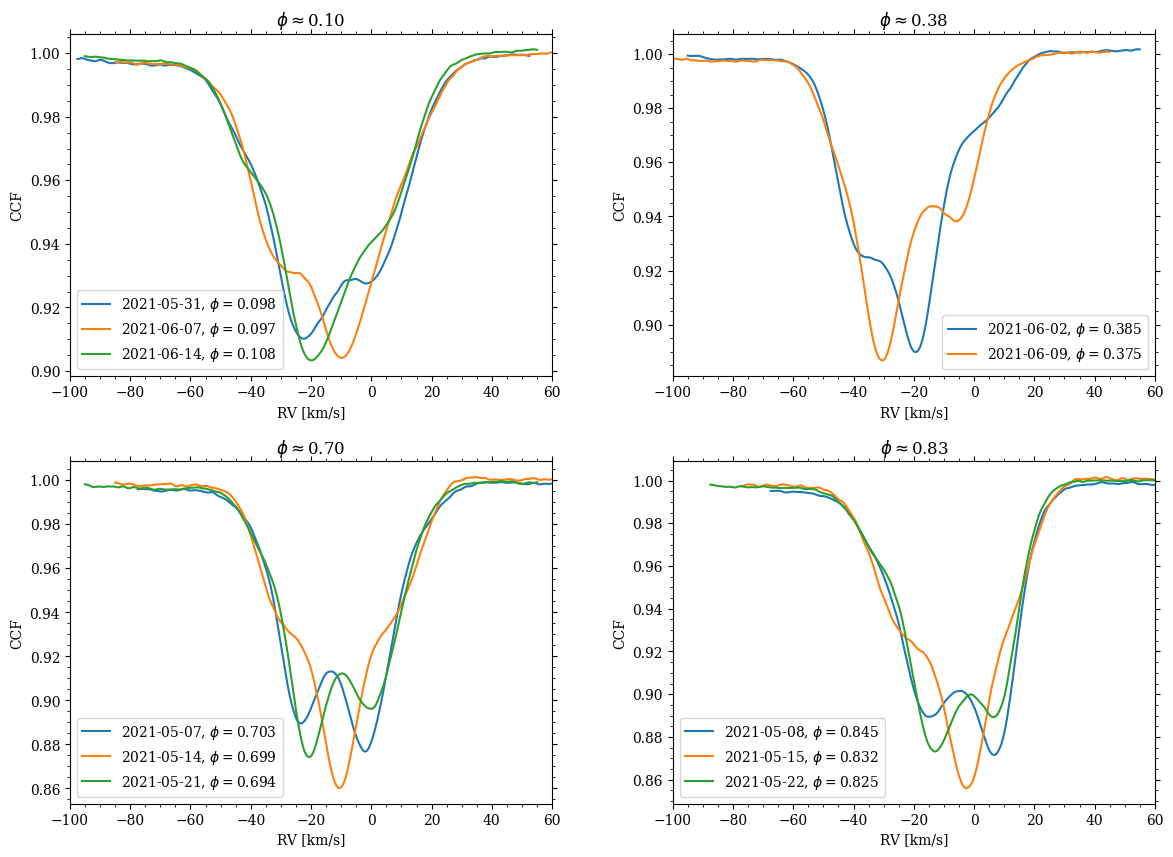}
    \caption{CCFs of X Sgr around the same phases of pulsation for consecutive cycles. Different line colors correspond to different BJD as indicated in the key. Phase of pulsation is marked on top of the panel.}
    \label{fig:xsgr_samephases}
\end{figure*}

\begin{figure}
    \centering
    \includegraphics[width=\linewidth]{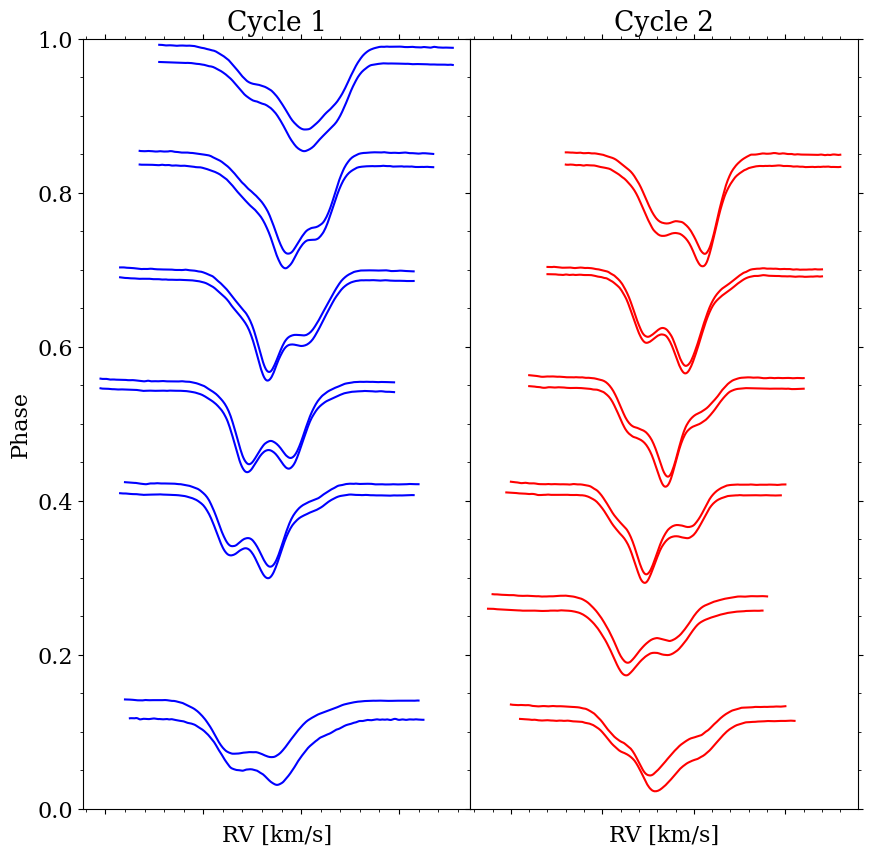}
    \caption{CCFs from two consecutive cycles (one cycle per panel) plotted according to pulsation phase for X Sgr.}
    \label{fig:xsgr_moreccfs}
\end{figure}

Line splitting in CCFs is the most prominent in the case of X Sgr among classical Cepheids studied in the literature and based on our sample of classical Cepheids observed by \veloce. Interestingly, in the case of X Sgr, line splitting is not limited to one phase of pulsation. In fact, \veloce\ data for X Sgr covers consecutive pulsation cycles thanks to the 7.01\,d pulsation period, which allows to investigate how the CCFs look like for similar phases of pulsation in consecutive cycles. This is presented in Fig~\ref{fig:xsgr_samephases} for phases $\phi \approx 0.10$, 0.38, 0.70, and 0.83 for at least two consecutive cycles. It is clear that CCFs from different cycles are significantly different from each other. The line splitting manifests differently even during the same pulsation phases. Hence, line splitting is neither limited to specific pulsation phases nor repeating on the timescale of the main pulsation mode as one would expect if it was connected to shock waves. In Fig.~\ref{fig:xsgr_moreccfs} we plotted CCFs from two full consecutive cycles to further show the evolution of CCFs.

\begin{figure}
    \centering
    \includegraphics[width=\linewidth]{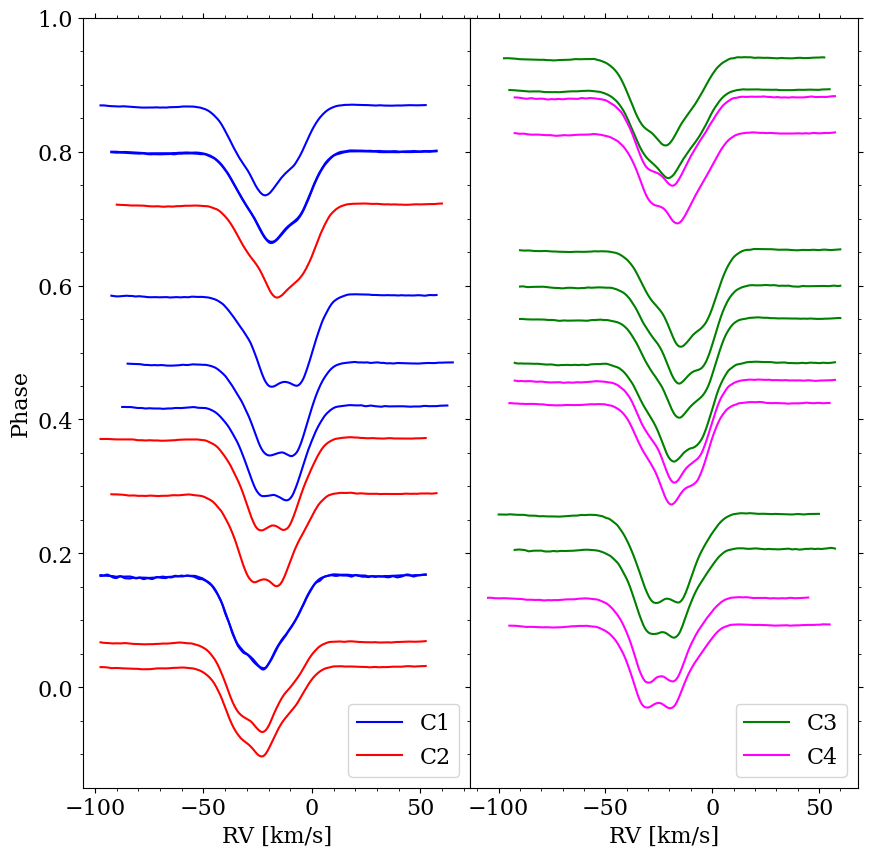}
    \caption{CCFs from four consecutive cycles (C1, C2, C3, and C4) plotted according to pulsation phase for BG Cru (two cycles per panel).}
    \label{fig:bgcru_moreccfs}
\end{figure}

In the case of BG Cru, we do not have data to plot the same phases of consecutive pulsation cycles. However, visual inspection of CCF profiles for similar phases of different cycles suggests, that this is also the case for BG Cru. Namely, that the line splitting observed in BG Cru is also not limited to specific pulsation phases but is periodic with a periodicity different than the main pulsation period. We plotted four consecutive cycles in Fig.~\ref{fig:bgcru_moreccfs}, two cycles per panel. While consecutive cycles show some resemblance, there are significant differences already two cycles apart.

\begin{figure*}
    \centering
    \includegraphics[width=\textwidth]{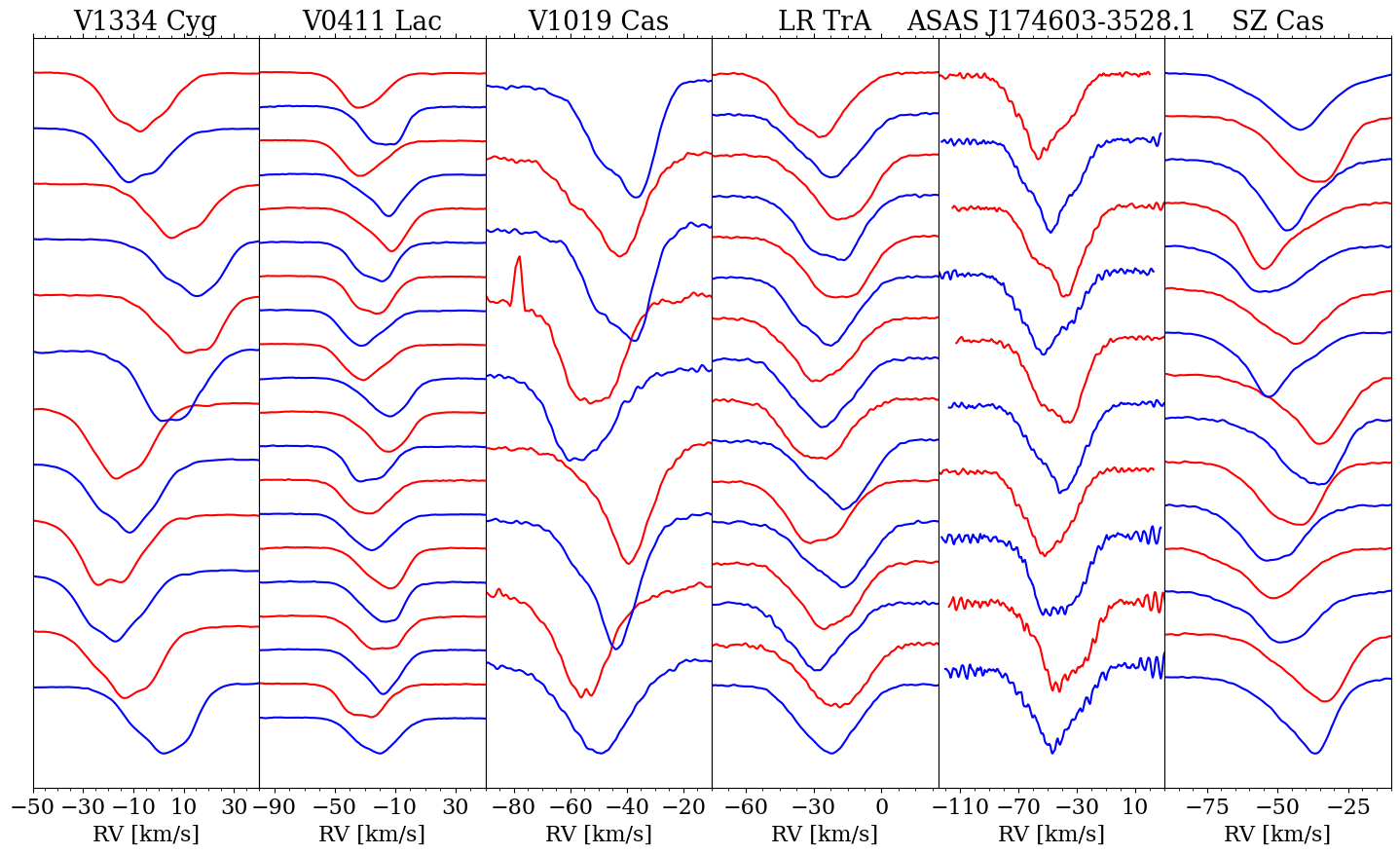}
    \caption{Selected CCFs of stars with humps/line splitting: V1334 Cyg, V0411 Lac, V1019 Cas, LR TrA, ASAS J174603-3528.1, and SZ Cas. Colors and vertical shifts are for better visualization.}
    \label{fig:ccfs_6stars}
\end{figure*}

\cite{anderson2013} reported line splitting in LR TrA, which we can confirm using the currently available \veloce\ data. Moreover, we found clear distortions of CCF profiles in five more Cepheids: V1334 Cyg, V0411 Lac, SZ Cas, V1019 Cas, and ASAS J174603-3528.1. In total, among 258 Cepheids from \veloce\ we found line splitting or unresolved line splitting in 8 stars, which gives the incidence rate of approximately 3 per cent. We plotted selected CCF profiles of LR TrA, V0411 Lac, V1334 Cyg, SZ Cas, V1019 Cas, and ASAS J174603-3528.1 in Fig.~\ref{fig:ccfs_6stars}, and we note that many of these stars are listed as modulators in \cite{veloce}. BG Cru and X Sgr are bright stars, with brightness in $V$ band of 5.47 mag and 4.58 mag, respectively. However, we detected similar features in stars as faint as 10 mag (for ASAS J174603-3528.1) or 11 mag (for V1019 Cas). Six out of eight stars with line splitting are first-overtone pulsators. Only two stars, X Sgr and SZ Cas, have dominant pulsations in the fundamental mode. 

\begin{figure*}
    \centering
    \includegraphics[width=\textwidth]{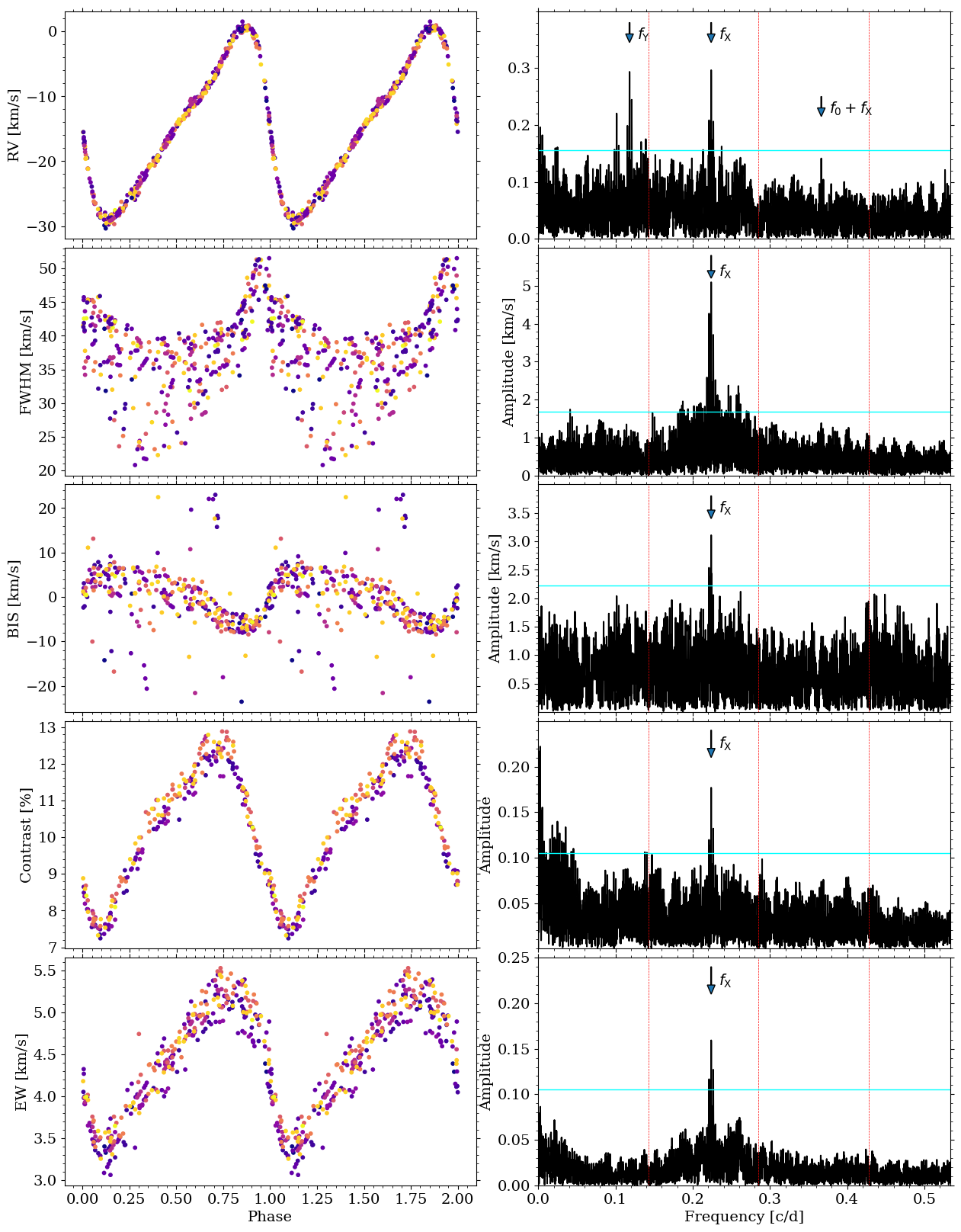}
    \caption{Left panel: data for X Sgr phased with the dominant pulsation period. Consecutive rows present data for RV, FWHM, BIS, contrast, and EW. BJD of each observation is color-coded. Right panel: Frequency spectrum after prewhitening with the dominant pulsation period and its harmonics. The position of the pulsation frequency and its harmonics are marked with red dotted lines. The additional signals are marked with arrows and labels (see also Table~\ref{tab:rv_frequencies}). The horizontal cyan line corresponds to three times the average noise level.}
    \label{fig:frequencygrams}
\end{figure*}

\begin{figure}
    \centering
    \includegraphics[width=\columnwidth]{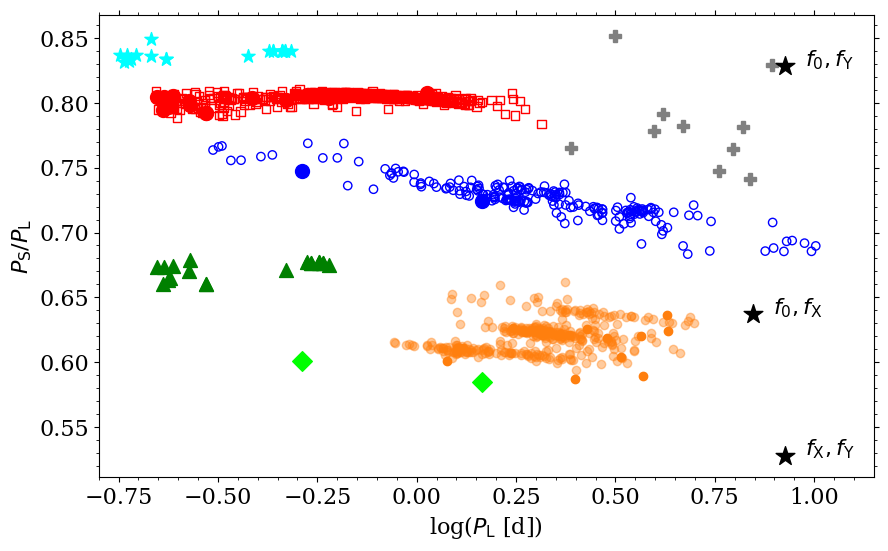}
    \caption{Petersen diagram for multi-mode classical Cepheids. Different types of multi-periodicity are plotted with different colors and symbols. Blue circles: pulsations in fundamental (F) mode and first overtone (1O). Red squares: pulsations in 1O and second overtone (2O). Cyan asterisks: pulsation in 2O and third overtone (3O). Green triangles: pulsations in 1O and 3O. Lime diamonds: pulsations in F and 2O. Orange points: 0.61 Cepheids. Grey pluses: subharmonics in 0.61 Cepheids. Data are combined from different sources, see references in \cite{netzel_review}. Signals detected in X Sgr are plotted with black asterisks and are labeled.}
    \label{fig:pet}
\end{figure}

\subsection{Frequency analysis of CCF shape indicators}\label{subsec:results_ccf_frequency_analysis}

As a first step of the frequency analysis, we used the RV provided by the \veloce\ project, as well as shape parameters: FWHM, BIS, EW, and contrast which were derived while measuring the RV for the \veloce. The RV, FWHM, BIS, contrast, and EW time-series of X Sgr phased with the dominant pulsation period and their corresponding frequency spectra after prewhitening with the dominant frequency, $f_0$, and its six harmonics are presented in Fig.~\ref{fig:frequencygrams}. Clearly, an additional signal, $f_{\rm X}$, is present in frequency spectra of all datasets. In the case of RV data, two additional signals are present. In Table~\ref{tab:rv_frequencies} we collected all signals detected in the frequency analysis of the RV data. Besides the two additional signals, $f_{\rm X}$, and $f_{\rm Y}$, we also detected a linear combination frequency, $f_{\rm X} + f_0$,  between $f_{\rm X}$ and the dominant periodicity, $f_0$. The $f_{\rm X}$ signal corresponds to the period of $P_{\rm X}=4.47$\,d, and forms a period ratio of around 0.637 with the dominant pulsation mode. 

X Sgr is plotted in the Petersen diagram (diagram of period ratio vs longer period) in Fig.~\ref{fig:pet} with multi-mode groups of classical Cepheids. Interestingly, the period ratio formed by the $f_{\rm X}$ with $f_0$ fits the long-period extension of the so-called 0.61 Cepheids, in which the additional signal forms a period ratio of around 0.61 -- 0.65 with the dominant first overtone (orange points in Fig.~\ref{fig:pet}; for a review see \citealt{netzel_review} and references therein). The $f_{\rm Y}$ signal would fit relatively well as a subharmonic of the $f_{\rm X}$ signal, i.e. $f_{\rm Y}/f_{\rm X} \approx 0.528$. However, the dominant pulsation mode of all currently known 0.61 Cepheids is the first overtone, while X Sgr primarily pulsates in fundamental mode (see RV curve characteristic for the fundamental mode in Fig.~\ref{fig:frequencygrams}; Fourier coefficients of the RV curve ($A_1 = 12.55$, $R_{21} = 0.4029$, $\phi_{21} = 3.1802$, $R_{31} = 0.1742$, $\phi_{31} = 0.2980$) are also consistent with values for the fundamental mode, see figures 2 and 3 of \citealt{hocde2023}), which is an argument against classifying X Sgr as a 0.61 Cepheid. The most promising hypothesis explaining the origin of the signals in the 0.61 Cepheids is due to harmonics of non-radial modes of degrees 7, 8, or 9 \citep{dziembowski2016}. However, the period ratio between harmonics of these non-radial modes and the fundamental mode should be different from 0.61. Consequently, the additional signals detected in X Sgr likely have a different origin than in the case of the 0.61 Cepheids, if the hypothesis behind the 0.61 Cepheids is correct.

Phased RV, FWHM, BIS, and contrast for BG Cru is published in fig.~1 in \cite{netzel_veloce}, where the additional signal forming a period ratio of around 0.61 was reported (see $P_X\approx 2.03$\,d, i.e. 0.49\,d$^{-1}$, in table 3 of \citealt{netzel_veloce}). BG Cru pulsates in the first overtone and was also classified by \cite{netzel_veloce} as a 0.61 Cepheid. The frequency analysis carried out by \cite{netzel_veloce} revealed yet another additional signal with frequency $f_{\rm Z} = 0.33235(3)$ d$^{-1}$, i.e. close to the first-overtone frequency of $f_{\rm 1O}=0.299$ d$^{-1}$. This signal is detectable in RV, FWHM, and BIS datasets. Interestingly, the $f_{\rm Z}$ is the only additional signal detected in the BIS periodogram. Its origin however was uncertain and not analyzed by \cite{netzel_veloce}. We analyze in Sec.~\ref{subsec:hump_freq} whether the $f_{\rm Z}$ signal can be connected to the observed line splitting. We note, that in addition to the analysis carried out in \cite{netzel_veloce}, we analyzed time series of EW. We did not detect any additional significant signals.

\begin{figure}
    \centering
    \includegraphics[width=\columnwidth]{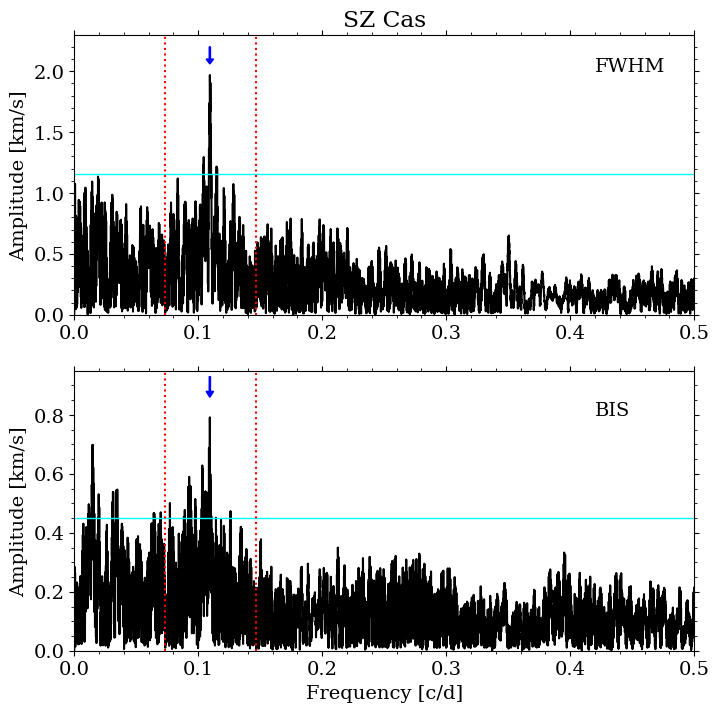}
    \caption{Frequency spectra for SZ Cas after prewhitening with the fundamental mode and its harmonic, which position is marked with red dotted lines. Additional signal is marked with the blue arrow. Top panel: frequency spectra calculated using FWHM time-series. Bottom panel: frequency spectra calculated using BIS time-series.}
    \label{fig:szcas_fs}
\end{figure}

We performed frequency analysis of RV, FWHM, BIS, and contrast data also for the six stars were unresolved line splitting was detected. In the case of SZ Cas, we detected additional signal in the frequency analysis of FWHM and BIS. Frequency spectra after prewhitening with the dominant fundamental mode and its harmonic are presented in Fig.~\ref{fig:szcas_fs}. SZ Cas fundamental-mode period is $P_0=13.638(1)$ d. The additional signal has a period of $P_{\rm X}=9.147(1)$. The shorter-to-longer period ratio is 0.67. The period and period ratio for SZ Cas does not correspond to any known groups of multi-mode classical Cepheids. Interestingly, this period ratio is relatively close to the position of subharmonic of the fundamental mode, i.e. $1.5f_0$. Still, it is unknown whether the additional periodicity is related to features observed in CCFs. Further observations are being collected to ascertain this. V0411 Lac was already analyzed by \cite{netzel_veloce} who reported a detection of an additional signal that has a period longer than the first-overtone period and both form a period ratio of around 0.687 (see discussion in Sec.~\ref{disc:occurence}). Besides X Sgr, BG Cru, SZ Cas, and V0411 Lac, we did not detect any additional signals in the rest of stars with humps.

\begin{table}
\caption{Signals found during the frequency analysis of RV data of X Sgr.}
    \centering
    \begin{tabular}{clll}
     Frequency &  $f$ [d$^{-1}$]   & A [km/s] & Phase [rad] \\
      \hline
     $f_0$	&	0.14260	&	12.55(3) &	5.132(4)	\\
2$f_0$	&	0.28519	&	5.05(3)	&	0.879(8)	\\
3$f_0$	&	0.42779	&	2.19(3)	&	3.13(2)	\\
4$f_0$	&	0.57039	&	0.76(3) &	5.03(4) \\
5$f_0$	&	0.71298	&	0.44(3)	&	0.70(6) \\
6$f_0$	&	0.85558	&	0.20(3)	&	3.1(1) \\
7$f_0$	&	0.99818	&	0.16(3)	&	5.0(2)\\
$f_{\rm X} $&	0.22382	&	0.31(3)	&	2.2(2)	\\
$f_{\rm Y}$	&	0.11815	&	0.31(3)	&	5.1(2)	\\
$f_0+f_{\rm X}$	&	0.36642	&	0.17(3) &	1.9(3)	\\
\\
    \end{tabular}
   \tablefoot{Consecutive columns provide signal's identification, its frequency, amplitude and phase.} 
    \label{tab:rv_frequencies}
\end{table}

\begin{figure}
    \centering
    \includegraphics[width=\columnwidth]{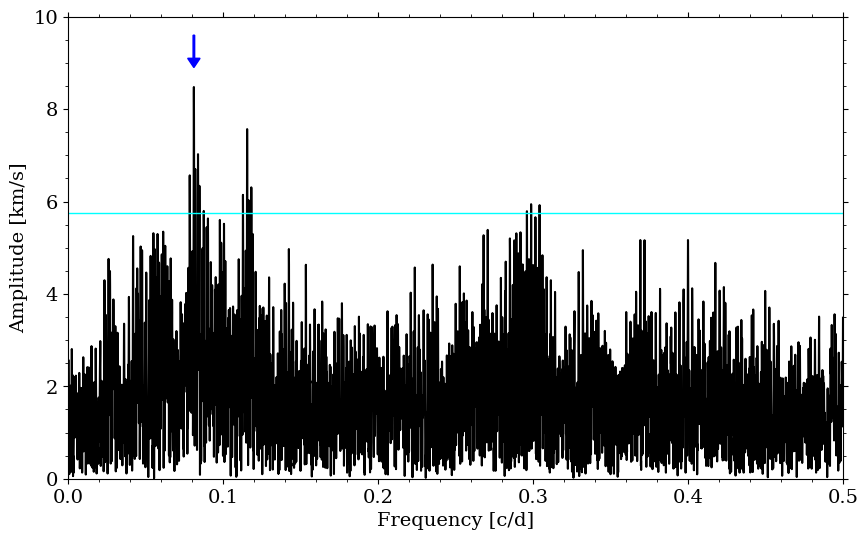}

    \includegraphics[width=\columnwidth]{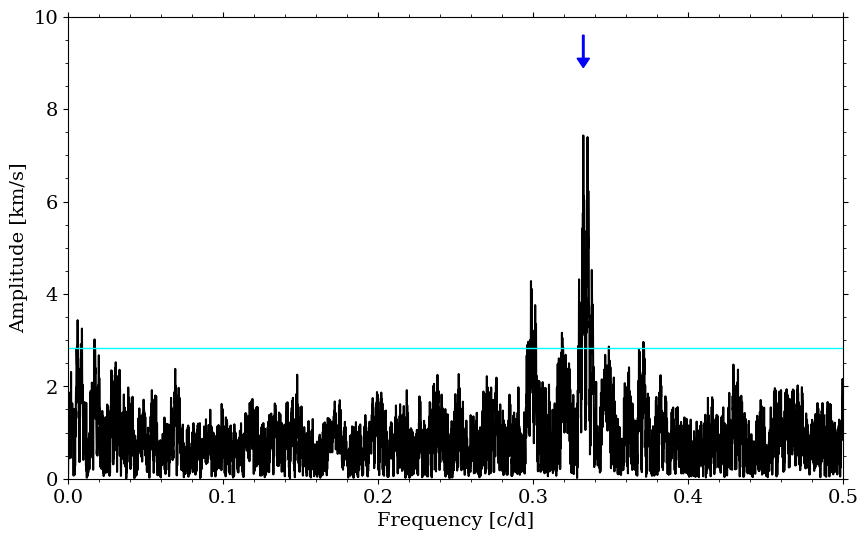}
    \caption{Frequency spectra for the time-series of RV of the hump in LPV (see text for details). The dominant signal is marked with the blue arrow. Top panel: frequency spectrum for X Sgr. Bottom panel: frequency spectrum for BG Cru.\label{increase label sizes, label the stars in the panels}}
    \label{fig:pypek_fs}
\end{figure}

\begin{figure}
    \centering
    \includegraphics[width=\columnwidth]{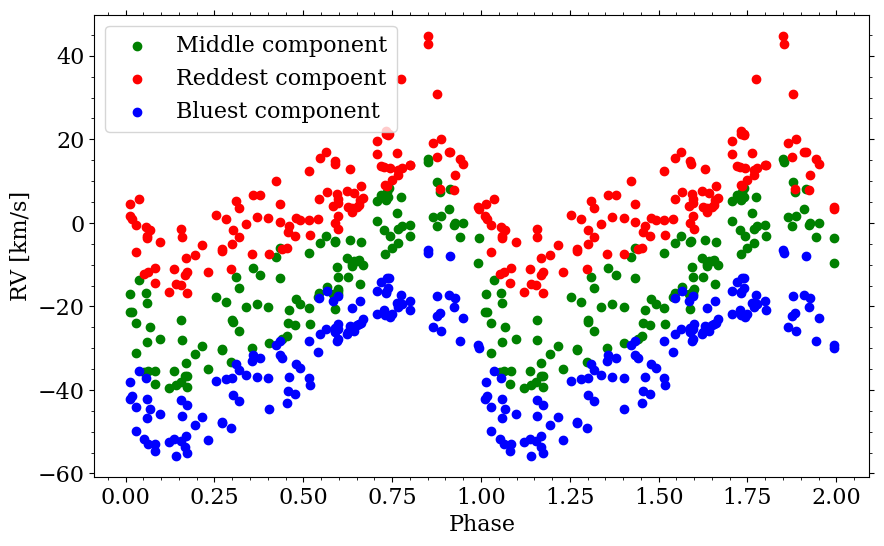}

    \includegraphics[width=\columnwidth]{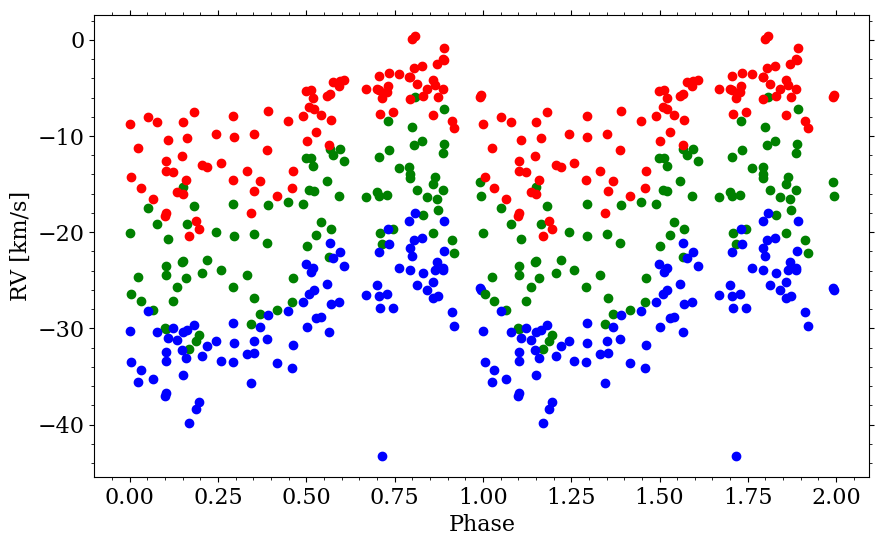}
    \caption{Centroids of three Gaussians used for the fit to CCF profiles phased with pulsation period for X Sgr (top panel) and BG Cru (bottom panel). Colors correspond to colors of the components in Fig.~\ref{fig:gaussian_fit}.}
    \label{fig:gaussian_centroid}
\end{figure}

\subsection{Frequency analysis of the hump feature}\label{subsec:hump_freq}

The CCFs and spectral lines of X Sgr and BG Cru exhibit humps or more partially separated components. We traced the evolution of the principal hump feature as described in Sec.~\ref{sec:methods} and shown in Fig.~\ref{fig:pypek}. Frequency spectrum of the relative hump position is presented in Fig.~\ref{fig:pypek_fs} for BG Cru and X Sgr. The dominant signal in the frequency spectrum of X Sgr is $f_{\rm X\,Sgr}=0.08190(3)$ d$^{-1}$ which corresponds to period of $P_{\rm X\,Sgr} = 12.317(4)$ d. Interestingly, the signal detected in frequency spectrum of RV, $f_{\rm X} = 0.22382$ d$^{-1}$, is very close to the combination frequency between the signal detected in frequency spectrum of a hump and dominant fundamental mode frequency, i.e. $f_0+f_{\rm X\,Sgr} = 0.22450$ d$^{-1}$. 

In the case of BG Cru, the highest-amplitude signal in the frequency spectrum of the hump time-series is $f_{\rm BG\,Cru}=0.33243(10)$ d$^{-1}$, which corresponds to the period of $P_{\rm BG\,Cru}=3.00813(9)$ d. Interestingly, the additional signal detected in frequency spectrum of RV, BIS and FWHM has $f_{Z} = 0.33235(3)$ d$^{-1}$ ($P_Z = 3.0088(3)$ d). Therefore the periodicity detected in the frequency spectra of RV, BIS and FWHM likely corresponds to the periodicity of the hump in CCF profiles.

\subsection{Triple-Gaussian fit results\label{sec:3G}}

\begin{figure}
    \centering
    \includegraphics[width=\columnwidth]{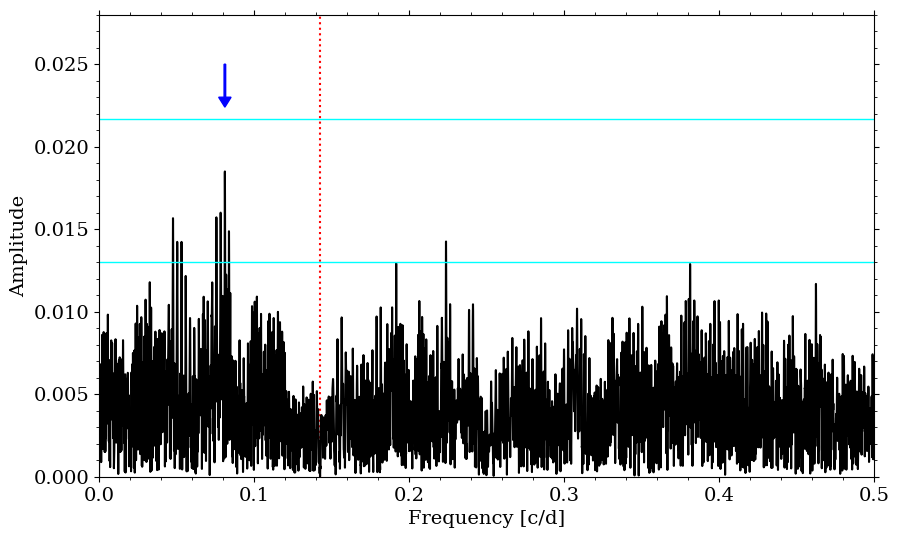}
    \caption{Frequency spectrum of a depth of the bluest Gaussian component (see Fig.~\ref{fig:gaussian_fit} and text for details) after prewhitening with the fundamental mode. Arrows marks the $f_{\rm X\,Sgr}$ signal. Red dotted lines marks the position of the fundamental mode frequency. Cyan lines corresponds to 3 and 5 times the average noise level.}
    \label{fig:ba_fs}
\end{figure}

Last, but not least, we analyzed the properties of the triple-Gaussian fits to CCFs. In the top panel of Fig.~\ref{fig:gaussian_centroid} we present centroids of the bluest, middle and the reddest Gaussian components from Fig.~\ref{fig:gaussian_fit} phased with the dominant pulsation period of X Sgr (top panel) and BG Cru (bottom panel). The data phase well with the dominant period, as expected. However, there is a significant additional scatter visible for all datasets. This scatter is significantly stronger for depth and width of the Gaussian components. Consequently, frequency analysis is hampered due to the relatively high noise level. The frequency analysis revealed the additional signals in some of the datasets, but only at $3\sigma$ level. The detected signals in the case of X Sgr are fundamental mode frequency, $f_{\rm X\,Sgr}$ (in depth of the reddest and bluest components, in centroids of all components, and in width of the bluest component; found previously in the analysis of the hump feature), $f_{\rm X}$ (in centroid of the middle component; found previously using line shape indicators), $f_0 + f_{\rm X}$ (in centroid of the blue component; found previously using line shape indicators), $f\approx 0.047$ d$^{-1}$ (in centroid of the blue component; not found previously). We note that some of these signals were detected only after prewhitening with the dominant pulsation mode. Moreover, none of these signals exceed the $5\sigma$ detection level. In the case of BG Cru we detected the first-overtone frequency, and $f_{\rm BG\,Cru}$ (in depth and centroid of the red and middle component; found previously in the analysis of the hump feature). Again, none of the additional signals exceeded the $5\sigma$ level. In Fig.~\ref{fig:ba_fs} we plotted the strongest detection of an additional signal in all of the datasets created from the triple-Gaussian fit, which is for the depth of the bluest Gaussian component for X Sgr.

\subsection{Fourier spectra with FAMIAS}

\begin{figure*}
    \centering
    \includegraphics[width=0.5\linewidth]{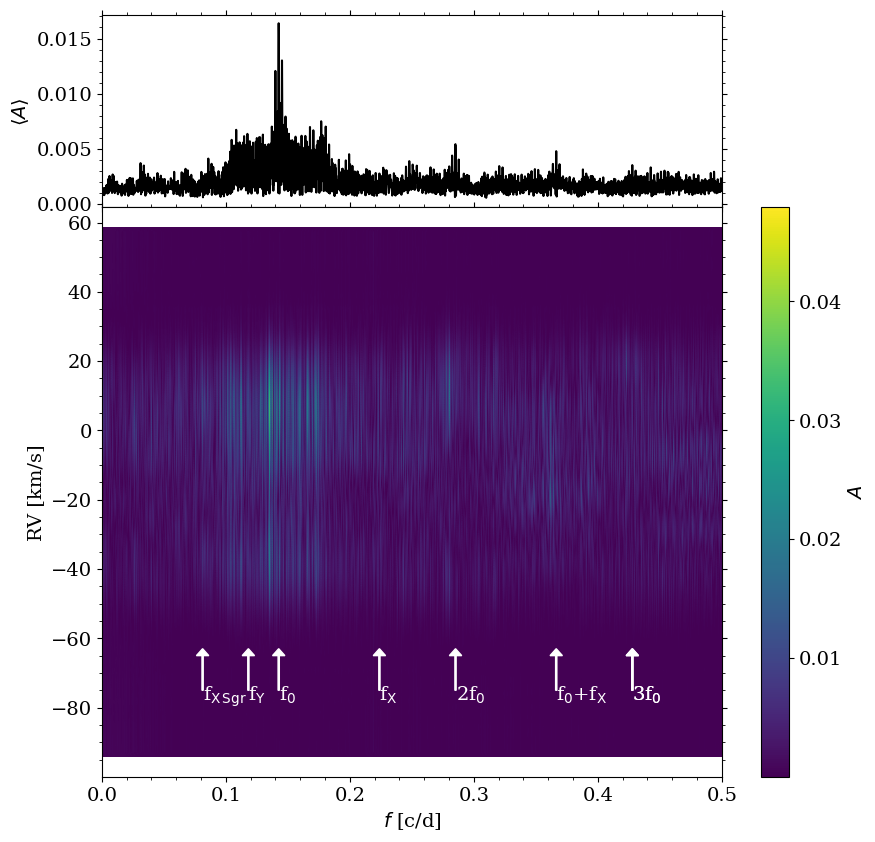}\includegraphics[width=0.5\linewidth]{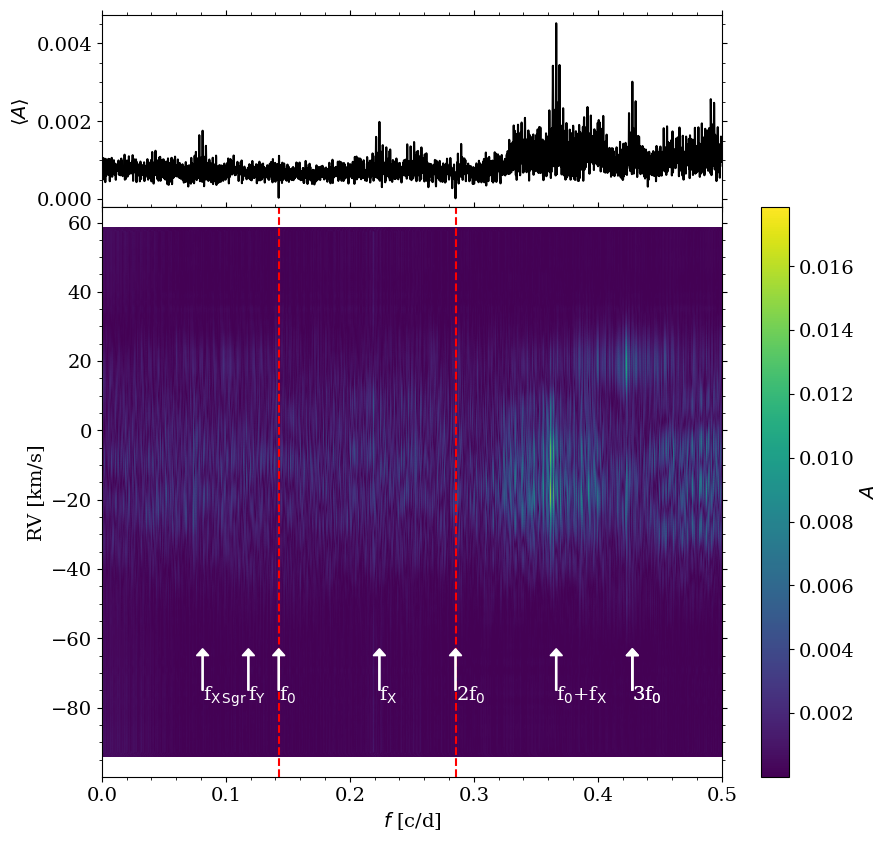}
    \caption{Fourier spectra in 1D and 2D for X Sgr calculated based on Coralie14 CCFs using FAMIAS software. Left panels: frequency spectra of the original data. Right panel: frequency spectra after prewhitening with the fundamental mode and its harmonic. Prewhitened frequencies are marked in the bottom panel with red lines. Bottom panels: 2D spectra, where amplitude is color coded. Frequencies found in other methods are marked with white arrows and labeled. Top panels: 1D mean Fourier spectra.}
    \label{fig:2d_1d_xsgr}
\end{figure*}

Two examples of 2D and 1D frequency spectra calculated for X Sgr with FAMIAS are presented in Fig.~\ref{fig:2d_1d_xsgr}. Signals detected using previous methods are marked with arrows. In the left panels of Fig.~\ref{fig:2d_1d_xsgr} we plotted frequency spectra of the original data. Clearly, the fundamental mode is visible. Also the harmonic of the fundamental mode is visible in the 1D mean Fourier spectrum, which is less noisy compared to 2D spectrum. However, based on the 2D spectrum, it is evident that the power is distributed asymmetrically across the CCF profile forming two asymmetric maxima. The same is barely visible for the harmonics of the fundamental mode. Interestingly, the first additional signal after prewhitening with the fundamental mode and its harmonic is the signal at around 0.366\,d$^{-1}$, which corresponds to the combination frequency $f_0+f_{\rm X}$ detected in the analysis of the RV data. Contrary to the radial mode, the highest amplitude is located at the center of the CCF. The list of frequencies detected using the 1D mean Fourier spectra is provided in Table~\ref{tab:xsgr_pbp}. The $f_{\rm Y}$ signal found in the analysis of the RV data was not detected with the 1D mean Fourier spectra. We included two interpretations for the detected signal. In interpretation A, we assumed that the independent signal is $f_{\rm X}=0.22$\,d$^{-1}$, as in the analysis of the RV data. Then the other signals are their combinations. In interpretation B we adopted the highest additional signal as the independent periodicity, i.e. $f_{\rm X}' = f_{\rm X} + f_0$. Unfortunately, it is not clear which scenario is correct and which signal is the independent periodicity. If the signals are due to non-radial modes, then it is possible for the combination frequencies to have higher amplitudes than the parent modes \citep[see][]{benko.kovacs2023, balona2013, kurtz2015}. Moreover, in the analysis of the hump position (see Sec.~\ref{subsec:hump_freq}), we found frequency of $f_{\rm X\,Sgr} = 0.08$\,d$^{-1}$, which was also detected in the 1D mean Fourier spectra as a combination frequency. Due to the fact that $f_{\rm X\,Sgr}$ was detected by directly tracing the hump position in the CCF profile, it is a good candidate for an independent periodicity, however, this idea would need another confirmation.

\begin{table}[]
\caption{Frequencies detected using prewhitening of the 1D mean Fourier spectra for Coralie14 CCFs of X Sgr.}
    \centering
    \begin{tabular}{llll}
    $f$ [d$^{-1}$] & $A$  & Interpretation A & Interpretation B \\
    \hline
    0.142581 &	2.51(5)	& $f_0$	& $f_0$	 \\
    0.285195 &	0.73(5)	& $2f_0$ & $2f_0$	\\
    0.366401 &	0.68(5)	 & $f_{\rm X}+f_0$ & $f_{\rm X}'$ \\
	0.42779 &	0.44(5)	 & $3f_0$ & $3f_0$ \\
	0.509015 &	0.42(5)	 & $2f_0+2f_{\rm X}$ & $f_0+f_{\rm X}'$ \\
	0.08123 &	0.27(5)	& $f_{\rm X\,Sgr}$ / $f_{\rm X} - f_0$ & $f_{\rm X}'-2f_0$ \\
	0.223825 &	0.27(5)	& $f_{\rm X}$  & $f_{\rm X}'-f_0$  \\
	0.570366 &	0.24(5)	 & $4f_0$  & $4f_0$   \\
  
         \\
    \end{tabular}
    \tablefoot{Consecutive columns provide frequency of the detected signal, its dimensionless amplitude and two alternative interpretations (see text for details).}
    \label{tab:xsgr_pbp}
\end{table}

\begin{table}[]
\caption{The same as in Tab.~\ref{tab:xsgr_pbp} but for BG Cru.}
    \centering
    \begin{tabular}{llll}
    $f$ [d$^{-1}$] & $A$  & Interpretation A & Interpretation B \\
    \hline
    0.299181 &	1.42(1) & $f_{\rm 1O}$	& $f_{\rm 1O}$	 \\
    0.598347 &	0.26(1)	& $2f_{\rm 1O}$ & $2f_{\rm 1O}$	\\
	0.631599 &	0.18(1)  & $f_{\rm 1O}+f_{\rm Z}$ & $f_{\rm Z}'$ \\
	0.03325 & 0.13(1)	 & $f_{\rm Z}+f_{\rm 1O}$  & $f_{\rm Z}'-2f_{\rm 1O}$  \\
	0.265912 &	0.10(1) & $2f_{\rm 1O}-f_{\rm Z}$ & $3f_{\rm 1O}-f_{\rm Z}'$ \\
	0.930782 &	0.10(1) & $2f_{\rm 1O}+f_{\rm Z}$   &  $f_{\rm 1O}+f_{\rm Z}'$  \\
    0.89753 & 0.07(1) & $3f_{\rm 1O}$ & $3f_{\rm 1O}$ \\
	0.565086 &	0.05(1)	 & $3f_{\rm 1O}-f_{\rm Z}$   &   $4f_{\rm 1O}-f_{\rm Z}'$   \\
    0.332441 & 0.05(1) & $f_{\rm Z}$ & $f_{\rm Z}'-f_{\rm 1O}$ \\
         \\
    \end{tabular}
   
    \label{tab:bgcru_pbp}
\end{table}

\section{Discussion}\label{sec:discussion}

\subsection{Line shape indicators in case of split profiles}

\begin{figure}
    \centering
    \includegraphics[width=\linewidth]{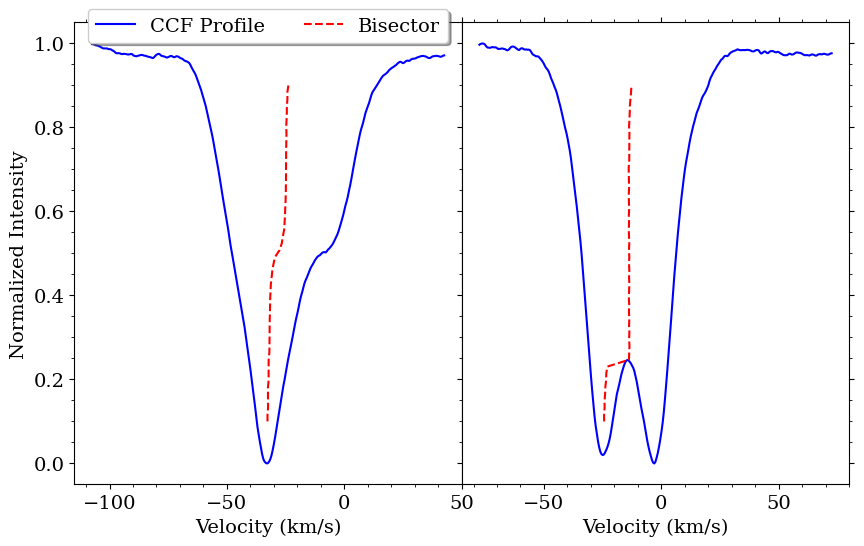}
    
    \includegraphics[width=\linewidth]{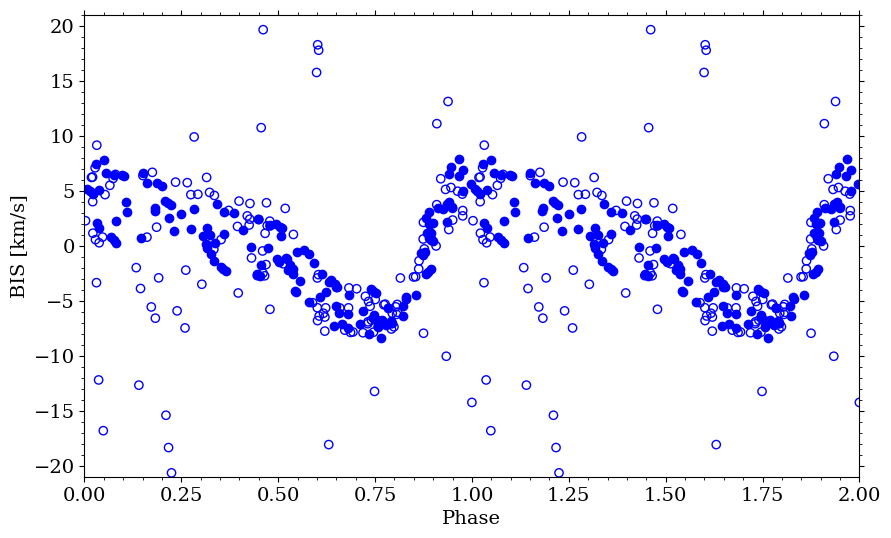}

    \includegraphics[width=\linewidth]{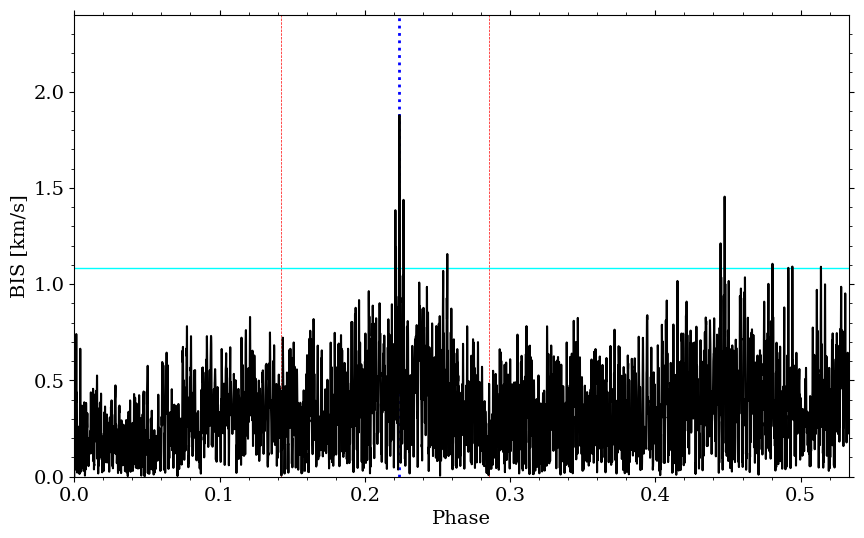}
    \caption{Bisector inverse span (BIS) for X Sgr. Top panel: visualization of BIS in the straightforward case (left) and in case of a split profile (right). Middle panel: Phased BIS data for straightforward cases (filled symbols) and for split profiles (open symbols). Bottom panel: Frequency spectrum after prewhitening with the dominant frequency and its harmonics calculated using only the straightforward cases.}
    \label{fig:good_bad_bis}
\end{figure}

Line shape indicators, in particular BIS, are well defined and easily calculated for CCF and line profiles of the majority of classical Cepheids. However, in the case of strongly deformed profiles as in the case of X Sgr, the definition of BIS is not as straightforward. We visualized this issue in the case of X Sgr in the top panel of Fig.~\ref{fig:good_bad_bis}. Namely, BIS is defined as midpoints of horizontal lines drawn across the CCF or line profile. In the case of a split profile, the bottom part of the profile consists of two dips so, the value of BIS is misrepresented as the line goes into one of the dips. To test whether using BIS for analysis in the case of X Sgr and BG Cru is substantiated, we visually inspected all calculated bisector lines for CCF profiles of X Sgr and flagged the cases when the profiles were split. Calculated BIS values phased with the fundamental period are plotted in the middle panel of Fig.~\ref{fig:good_bad_bis}, where the filled symbols correspond to the profiles where BIS could be easily calculated and open symbols are split profiles. The vast majority of points corresponding to the split profiles are outliers (many of them are not limited to the range shown in the phase plot) and do not phase well with the fundamental period. BIS values calculated from the regular profiles phase well with the fundamental period. Hence, the apparent noisiness of BIS in X Sgr is a consequence of line splitting, not of poor S/N of the CCFs. Note, that in the case of BG Cru there are fewer such outliers in BG Cru where the line splitting is less extreme. The frequency spectrum for X Sgr calculated using only the verified values of BIS is presented in the bottom panel of Fig.~\ref{fig:good_bad_bis} and clearly shows the same additional signal as in Fig.~\ref{fig:frequencygrams}. Removing the outliers connected to the split profiles lowered the overall noise level. However, even with the outliers caused by the split profiles, the BIS is proven to be a good dataset to search for additional signals.

\subsection{On the occurrence of line splitting in CCFs\label{disc:occurence}}

\begin{figure}
    \centering
    \includegraphics[width=\columnwidth]{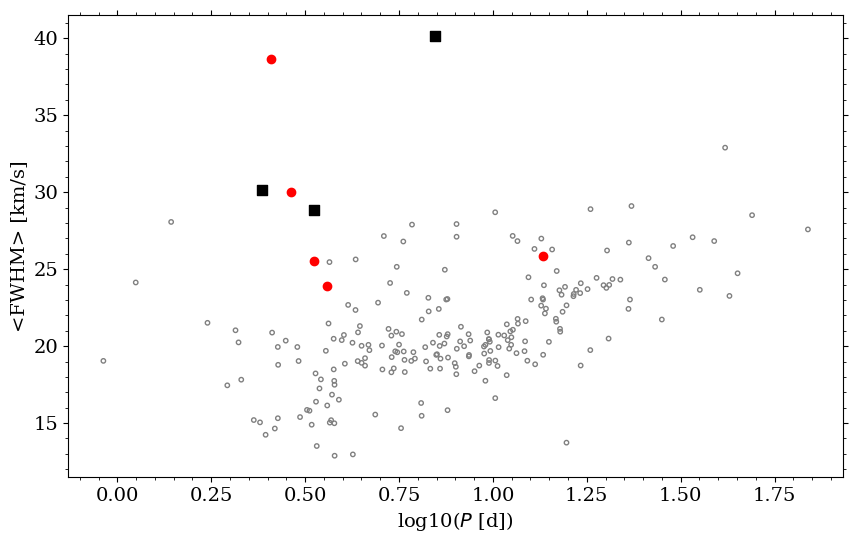}
    \caption{Average FWHM calculated from all Gaussian fits to CCFs of classical Cepheids observed by \veloce as a function of the dominant pulsation period. Stars with regular CCF profiles are plotted with grey circles. X Sgr, BG Cru, and LR TrA are plotted with black squares. Additional selected candidates showing humps and line splitting in CCFs are plotted with red points.}
    \label{fig:fwhm_period}
\end{figure}

We found line splitting or unresolved line splitting in eight stars out of 285 Cepheids monitored by \veloce. Interestingly, the eight stars where line splitting or humps are unambiguously detected have relatively high values of average FWHM compared to the rest of classical Cepheids observed by \veloce. This is presented in Fig.~\ref{fig:fwhm_period}, where the average FWHM is plotted against the dominant pulsation period. The record holder is X Sgr, where average FWHM is 40 km/s, while BG Cru and LR TrA have values around 30 km/s. ASAS J174603-3528.1 is another star with high average FWHM of 39 km/s. While its CCFs do not have the same quality as for X Sgr, the humps in CCFs profiles are clearly visible (see the fifth panel of Fig.~\ref{fig:ccfs_6stars}). This preference for higher values of average FWHM than typical suggests that line splitting might be correlated with high FWHM. It is not clear whether line splitting is connected to faster rotation manifesting as larger FHWM values, or if wider profiles make line splitting easier to detect. 24 Cepheids from our sample have literature values of $v\sin i$, including X Sgr and BG Cru \citep{xsgr_rot}. The literature values of $v\sin i$ cover a wide range from 3.3 km/s for MY Pup to 26 km/s for X Sgr, with a typical value of around 15 km/s for the sample. Interestingly, BG Cru also has one of the highest values of $v\sin i = 21$ km/s.

\begin{figure}
    \centering
    \includegraphics[width=\columnwidth]{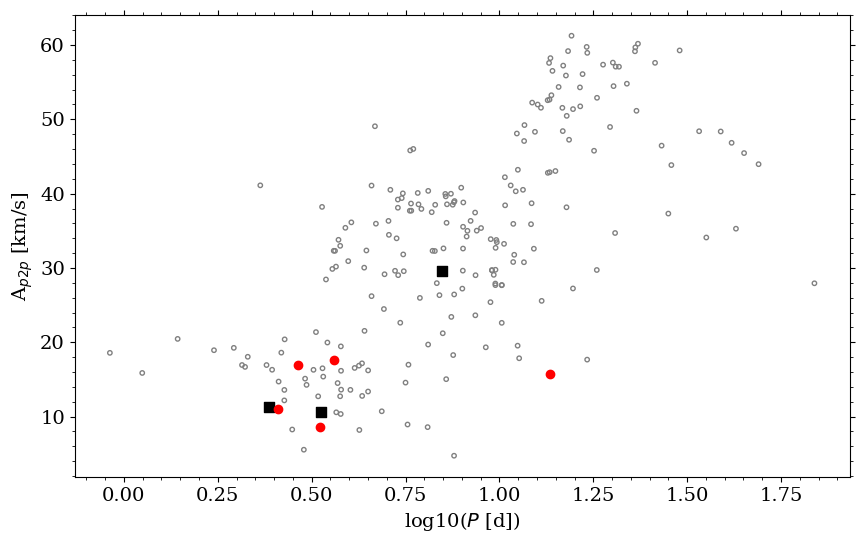}
    \caption{Peak-to-peak amplitude of RV curve as a function of the dominant pulsation period for all classical Cepheids from \veloce\ and stars with (unresolved) line splitting. Meaning of symbols the same as in Fig.~\ref{fig:fwhm_period}.}
    \label{fig:ap2p_period}
\end{figure}

\begin{figure}
    \centering
    \includegraphics[width=\columnwidth]{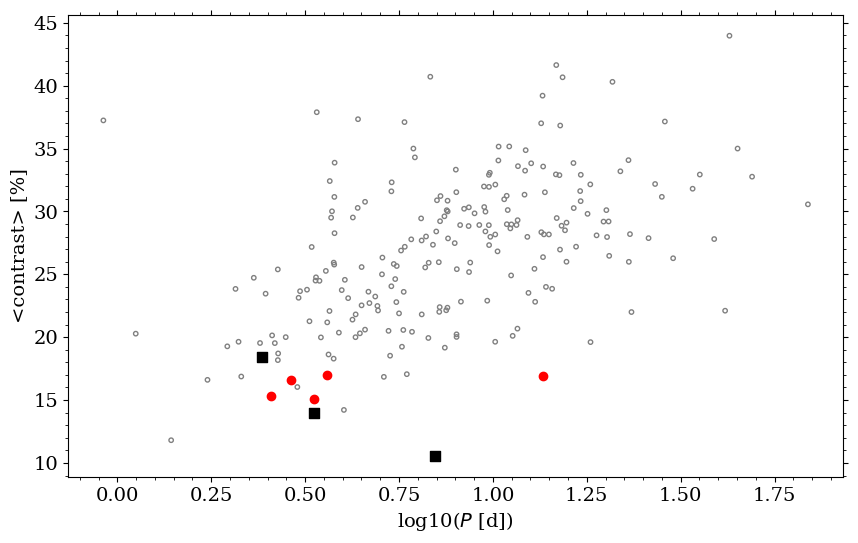}
    \caption{Average contrast as a function of the dominant pulsation period for all classical Cepheids from \veloce\ and stars with (unresolved) line splitting. Meaning of symbols the same as in Fig.~\ref{fig:fwhm_period}.}
    \label{fig:depth_period}
\end{figure}

Additionally, these eight stars tend to have peak-to-peak amplitude of the RV curve from a lower half of the range defined by all 258 \veloce\ Cepheids. Peak-to-peak amplitudes for all Cepheids and eight stars with line splitting or humps are plotted in Fig.~\ref{fig:ap2p_period}. The highest values for the observed Cepheids is up to 61 km/s. None of the eight stars with (unresolved) line splitting have a higher amplitude than 29 km/s. Average depth for the sample is plotted in Fig.~\ref{fig:depth_period}. Stars with (unresolved) line splitting have lower values than the rest of the sample. X Sgr has the lowest value among them -- 10.6 percent. Moreover, none of the stars with humps or line splitting have an average depth higher than 20 percent.

Two stars are particularly interesting: V0411 Lac and SZ Cas. V0411 Lac was already analyzed by \cite{netzel_veloce} who performed a frequency analysis of CCF shape indicators. The authors reported a detection of another low-amplitude additional signal that has a longer period than the first overtone and forms a period ratio of around 0.687 with the first overtone. This period ratio was already reported in the literature in some of first-overtone classical Cepheids based on photometric observations, but the origin of the signal remains unkown \citep[see a discussion in][and references therein]{netzel_veloce}. Here, we performed a frequency analysis for SZ Cas. The additional signal was detected in the FWHM and BIS data (see Sec.~\ref{sec:results}). The additional periodicity forms a period ratio of 0.67 with the main pulsation period. However, opposite to V0411 Lac, the additional signal has a period shorter than the main period. Unfortunately, the dataset currently available for V0411 Lac and SZ Cas is not numerous enough for analysis similar to X Sgr and BG Cru. Hence, we cannot confirm whether the periodicity detected by \cite{netzel_veloce} in V0411 Lac and the signal detected in SZ Cas in this work are related to the observed humps in CCFs.

\subsection{Interpretation of detected signals}

In both X Sgr and BG Cru, we detected additional signals in frequency spectra of RV and CCF profiles shape indicators. Frequency analysis of photometric data for X Sgr collected by BRITE revealed additional signals as well \citep{smolec_brite}, which are equivalents of $f_{\rm Y}=0.11815$ d$^{-1}$ and $f_{\rm X}=0.22382$ d$^{-1}$ detected in time-series of RV, FWHM, BIS, EW, and contrast data of \veloce. Possible hypotheses considered by \cite{smolec_brite} include pulsations in non-radial modes, periodic modulation or contamination. Contamination scenario is unlikely, since we also independently detected these signals together with combination frequencies. We also did not find any star using the Gaia DR3 catalog \citep{gaiadr3} within 2 arcsec of X Sgr, where this distance corresponds to the fiber size. Note however, that X Sgr is bright, so detection of any close companion would be significantly hampered. On the other hand, for a contaminating star to affect spectra, it would need to be considerably bright. In that case, such star would be expected to affect photometric light curves significantly as well which is not the case. Consequently, the scenario of contamination is unlikely given the fact that the same signals are detected in both photometric and spectroscopic dataset. We also did not notice any clear signature of long-term periodic modulation based on over ten years of the \veloce\ data. From the interpretation of \cite{smolec_brite} we are left with non-radial modes. Moreover, \cite{kovtyukh2003} also detected unexpected features in line profiles of X Sgr and BG Cru and proposed that the origin behind them are pulsations in non-radial modes.

Note, that in the case of BG Cru, there are multiple additional signals detected based on the frequency analysis of the CCF shape indicators \citep{netzel_veloce}. One of them, $f_{\rm Z} = 0.33235(3)$ d$^{-1}$, corresponds exactly to the periodicity obtained through the analysis of the hump feature in the CCFs. Hence, we conclude that the detection of $f_{\rm Z}$ in frequency spectra of RV, FWHM, and BIS is directly related to the line splitting phenomenon visible in CCFs. However, $f_{\rm Z}$ is not the highest-amplitude additional signal in CCF shape indicators frequency spectra. \cite{netzel_veloce} already reported an additional signal in BG Cru detected based on the frequency analysis of the FWHM and BIS data that has a period of around 2.03\,d. This period forms a period ratio of around 0.61 and was already connected to likely pulsation in non-radial modes of moderate degrees of $\ell = 7, 8, 9$ \citep{dziembowski2016}. Hence, $f_{\rm Z}$ would be another non-radial mode present in BG Cru, if the interpretation of $f_{\rm Z}$ is indeed due to non-radial mode. It is not clear however, which non-radial mode can be at play.

On the other hand, \cite{mathias2006} analyzed spectra of X Sgr and proposed that the observed line splitting features in line profiles are due to pulsation-induced shock waves. In fact, the authors noticed that line profiles can be reproduced with not two but three components \citep[see fig. 1 in][]{mathias2006}. Indeed shock waves manifest in line profiles as line splitting as known in other types of pulsating stars such as Miras or RR Lyrae stars \citep{alvarez2001, fokin.gillet1994}. However, pulsation-induced shock waves should be limited to specific phases of pulsation. This is not the case of X Sgr since its line/CCF profiles appearance is significantly different for virtually the same pulsation phase (see Fig.~\ref{fig:xsgr_samephases}). Moreover, we showed that changes in the hump feature, which appears when the profile is split, in X Sgr has indeed an underlying periodicity which is significantly different than the dominant pulsation period. Consequently, both observations effectively rule out the possibility that this phenomenon is caused by the pulsation-induced shock waves.

The periodicity of the hump in CCFs may in principle arise also as a result of a feature on the surface of a rotating star. Interestingly, indeed tracing the relative position of the hump reveals a periodic phenomenon. Moreover, it appears that line splitting may correlate with the FWHM of the CCFs (see Sec.~\ref{disc:occurence}), indicating a possible link to rotation. Assuming the period-radius relations from \citet{Anderson2016rot}, BG Cru and X Sgr would have radii of approximately $41$ and $53\,R_\odot$, respectively. For comparison, \citet{LiCausi2013} estimated $R = 53 \pm 3\,R_\odot$ for X Sgr using long-baseline interferometric observations. Assuming a rotation period of $3\,$d for BG Cru would thus yield a very fast surface velocity of $110\,\mathrm{km\,s^{-1}}$, while for X~Sgr, $f_X$ would correspond to $v_{\rm eq} = 96\,\mathrm{km\,s^{-1}}$ and $f_1$ to $v_{\rm eq} = 35.8\,\mathrm{km\,s^{-1}}$. For comparison, the minimum FWHM of X Sgr's CCFs is $\sim 34.1\,\mathrm{km\,s^{-1}}$, and $27.8\,\mathrm{km\,s^{-1}}$ for BG Cru, implying inclinations $i \lesssim 15-20$\,degrees for both stars. Assuming a mass of $\sim 5$, a spherical star of $50\,R_\odot$ would have a first critical velocity ($v_c=\sqrt{GM/R}$) of $\sim 140\,\mathrm{km\,s^{-1}}$. However, the predicted rotation rates of blue loop Cepheids formed from very fast rotating main sequence stars are much lower than this, typically between $20-30\,\mathrm{km\,s^{-1}}$ \citep[see tab. A.1 and A.4 in][]{Anderson2016rot}. Although an interpretation of $f_X$ in terms of surface rotation would yield a value smaller than $v_c$, it would nonetheless be a factor of $3-4$ larger than predicted by single-star evolution models, thus requiring to invoke binary interactions. Neither X~Sgr nor BG~Cru are spectroscopic binaries \citep{veloce}.

\begin{figure*}
    \centering
    \includegraphics[width=\linewidth]{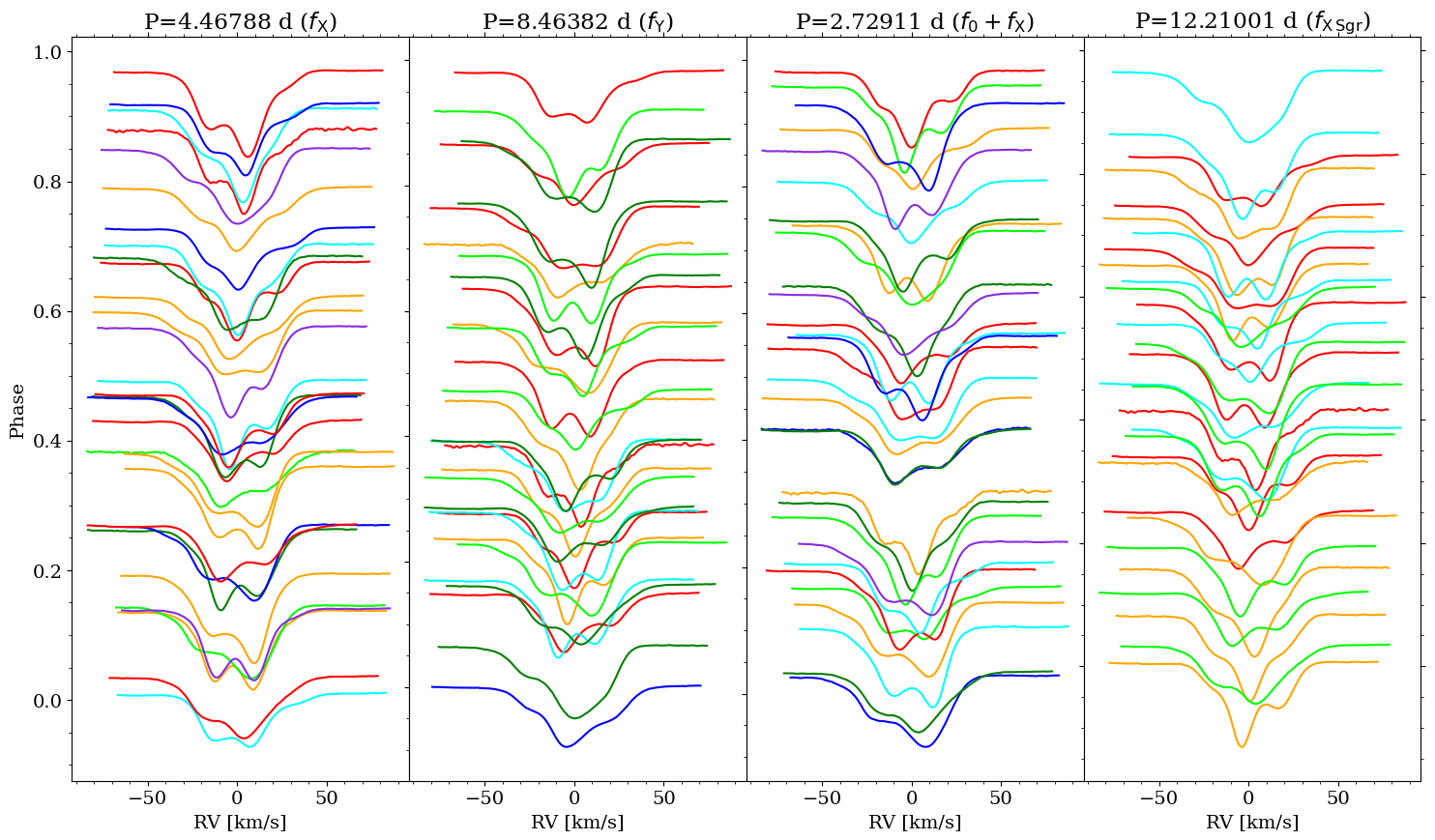}
    \caption{CCFs of X Sgr phased based on four different periods indicated at the top of each panel. All CCFs are shifted according to RV for each observation. CCFs plotted here were collected for BJD from 2459338 to 2459383. Colors differentiate between different cycles.}
    \label{fig:xsgr_ccf_vgamma}
\end{figure*}

In Fig.~\ref{fig:xsgr_samephases} we showed that CCFs of X Sgr look strikingly different for the same pulsation phases. In addition to this, we showed in Fig.~\ref{fig:xsgr_ccf_vgamma} how CCFs look like for the same phases when they are phased according to different additional signals that we found during our analysis. We considered $f_{\rm X}$ and $f_{\rm Y}$ from the analysis of the shape indicators, the combination signal $f_{\rm 0} + f_{\rm X}$, and $f_{\rm X\,Sgr}$ from the hump tracing analysis. For better visualization, all CCFs were shifted by the RV of each observation. It is clear, that the CCFs (shifted to the mean velocity of each star) for similar phases are significantly different, regardless of the period used to calculate the phase. This shows that the changes in CCFs are either not strictly periodic or there are multiple periods at play. 

\begin{figure}
    \centering
    \includegraphics[width=\linewidth]{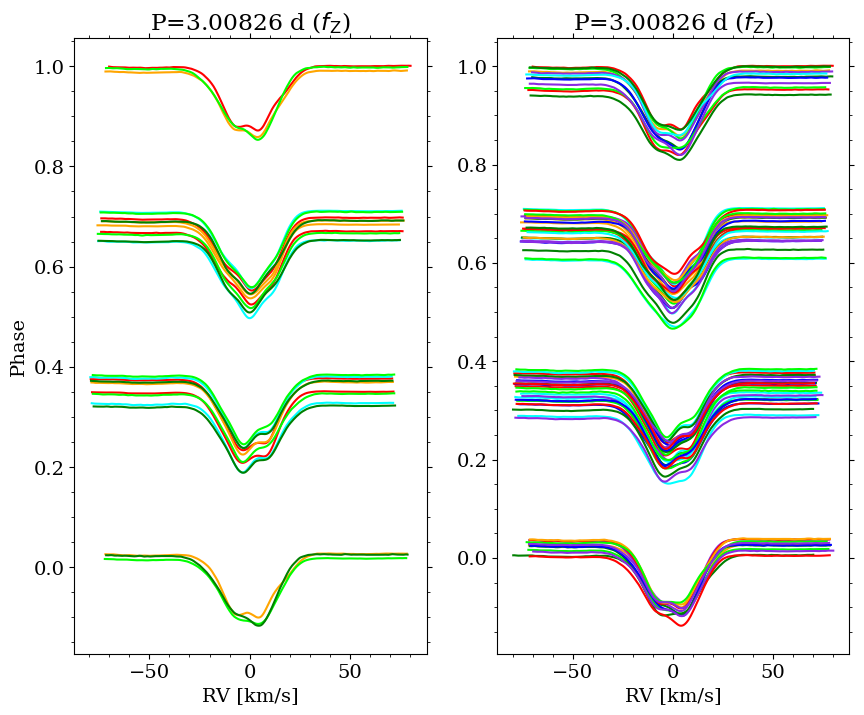}
    \caption{CCFs of BG Cru phased based on the period found in the hump analysis. All CCFs are shifted according to RV for each observation. Colors differentiate between different cycles. Left panel: CCFs were collected for BJD from 2459974 to 2459989 (five cycles of $P_{\rm Z}$). Right panel: CCFs were collected for BJD from 2459974 to 2460019 (fifteen cycles of $P_{\rm Z}$).}
    \label{fig:bgcru_ccf_vgamma}
\end{figure}

In Fig.~\ref{fig:bgcru_ccf_vgamma} we plot CCFs of BG Cru, phased according to the period $P_{\rm BG\,Cru}$ from the hump analysis, which was also independently found as $P_{\rm Z}$ based on the CCF shape indicators. CCFs were also shifted according the the RV for each observation. On the two different panels we plotted CCFs collected for five and fifteen consecutive cycles of the additional periodicity. Clearly, the CCFs look similar for similar phases when we consider five cycles. On the other hand, when considering fifteen consecutive cycles, there are some differences between CCFs arising. 

In BG Cru and X Sgr the line splitting phenomenon is very strong. Moreover, the data collected for BG Cru and X Sgr in \veloce\ is numerous and have high signal-to-noise since both stars are bright. Therefore, we were able to carry out  a detailed analysis of the line splitting phenomenon in both stars resulting in detection of the underlying periodicities. We note however, that line splitting was also reported in several other stars in the literature. In particular, it was also reported in LR TrA by \cite{anderson2013}. However, the dataset available in \veloce\ for LR~TrA is not numerous enough, given that the line splitting is not as strong as in the case of X Sgr or BG Cru, to perform the similar frequency analysis. The \veloce\ observations are ongoing and we aim to repeat this analysis for LR TrA in the upcoming years.

\cite{kovtyukh2003} also reported unexpected line profile variations in EV Sct and V1334 Cyg. The latter star is also a part of the \veloce\ database and  based on our dataset we also confirm the distortions of the CCF profiles (see Fig.~\ref{fig:ccfs_6stars}). Unfortunately, the detailed analysis of the hump periodicity similar to that presented for X Sgr and BG Cru is not possible at present in the case of V1334 Cyg. More observations are being carried out in order to enable similar detailed analysis.

\subsection{Line splitting in spectral lines}

CCFs benefit from very high signal-to-noise, allowing for a detailed analysis of their shapes and variability. However, they come at a loss of information from individual lines. Interestingly, it was already reported that line splitting in metallic lines in X Sgr does not manifest in Balmer lines \citep{mathias2006}. Using our spectra, we also confirm this for X Sgr. 

\begin{figure}
    \centering
    \includegraphics[width=\columnwidth]{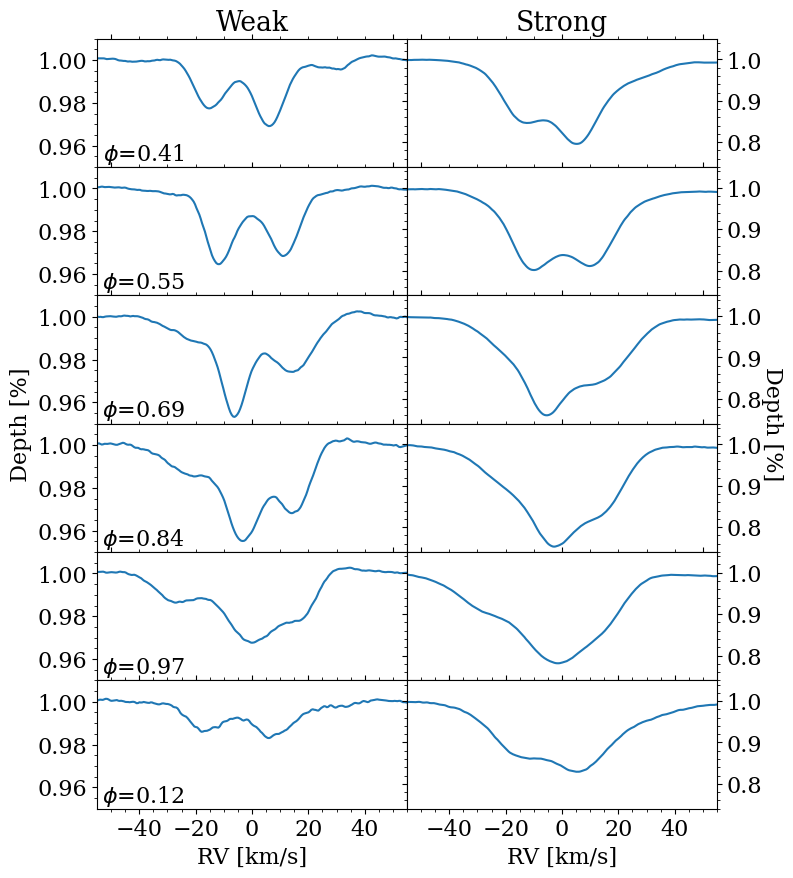}
    \caption{Examples of CCFs for X Sgr calculated using the weak line mask (left) and strong line mask (right) for the same pulsation phases on one pulsation cycle. The pulsation phase is indicated in the bottom left corner of left panels. CCFs are shifted using the mean RV for better visualization of the shape variations.}
    \label{fig:low_high_mask}
\end{figure}

Individual metallic lines are subject to relatively high noise. To compromise between investigating individual lines and benefiting from high signal-to-noise provided by the CCFs, we constructed two additional masks to calculate two additional sets of CCFs. The two additional masks were created based on the original G2 mask by selecting weak (depth < 0.55) and strong (depth > 0.65) lines. These masks were already used to investigate cycle-to-cycle modulations in long-period Cepheids by \cite{anderson2016_masks}. In Fig.~\ref{fig:low_high_mask} we plotted CCFs based on the two masks for the same pulsation phases for X Sgr. Note, that there is a difference in depth between weak and strong CCFs. CCFs based on weak lines are significantly more distorted than those calculated from strong lines. In particular, for phases $\phi=0.84$ and $\phi=0.97$ of the presented pulsation cycle, three components are clearly visible in the case of weak CCF, while only mild distortion is present in the case of strong CCF. 

Weak metallic lines differ from the strong lines by their low and high excitation potential. Consequently, weak metallic lines are more strongly constrained by the location where they are formed. On the contrary, strong lines are formed at a larger range of optical depths. Hence, the line splitting is likely getting averaged out over a larger part of an atmosphere than in the case of weak lines.

\section{Conclusions}\label{sec:conculsion}

We analyzed in detail line splitting present in BG Cru and X Sgr using CCFs collected by the \veloce\ project \citep{veloce}. We also searched for more stars showing line splitting or humps in CCFs indicating the same phenomenon using all 258 classical Cepheids observed within the \veloce\ project. Our findings and conclusions are following:

\begin{itemize}
\item We showed that CCFs differ significantly for the same pulsation phases of consecutive cycles for X Sgr. This rules out the possibility that the observed line splitting is connected to specific pulsation phases.
\item We traced the evolution of the hump formed by the line splitting by calculating its relative radial velocity compared to the radial velocity of the Gaussian fit to the CCF. Frequency analysis of such time-series revealed a periodicity of around 12\,d for X Sgr. 
\item The same analysis was performed for BG Cru and revealed the periodicity of the hump in CCFs of around 3\,d. The same periodicity was detected in frequency spectra of CCF shape indicators (RV, FWHM, BIS) besides the dominant frequency of the radial mode, its harmonics, and the additional signal forming period ratio of around 0.61.
\item We performed frequency analysis of CCF shape indicators for X Sgr, which revealed additional periodicity of 4.47\,d, and (only for RV) of 8.46\,d.  
\item We detected similar CCF features in six more stars: LR TrA, SZ Cas, V0411 Lac, V1334 Cyg, V1019 Cas, and ASAS J174603-3528.1. In total, this gives 8 stars with humps/line splitting selected from 258 classical Cepheids, which corresponds to the incidence rate of 3 per cent. Among the 8 stars, 2 are fundamental-mode (SZ Cas, X Sgr), and remaining 6 are first-overtone pulsators.
\item Stars with humps/line splitting tend to have higher average FWHM than typical for classical Cepheids observed by the \veloce\ project. The record holder is X Sgr, with 40 km/s, where the line splitting phenomenon is the strongest.
\item Stars with humps/line splitting tend also to have peak-to-peak RV amplitudes below 30 km/s, i.e. they have values from the lower half of the possible range defined by the \veloce\ sample.
\item Stars with humps/line splitting also have an average contrast lower, below 20 percent, i.e. from the lower end of the distribution of the whole sample. The record holder is X Sgr with just 10.6 percent.
\item Four stars out of eight show additional signals in frequency spectra of CCF shape indicators: BG Cru, X Sgr, V0411 Lac and SZ Cas.
\item V0411 Lac has an additional long-period signal forming period ratio of 0.687 with the dominant first overtone. This star is a member of a group of multi-mode Cepheids showing this period ratio that were identified based on photometric observations.
\item SZ Cas show an additional shorter-period low-amplitude signal that forms a period ratio of around 0.67. The origin of this periodicity is unknown.
\item We investigated how line splitting manifests in strong and weak metallic lines by using modified masks to calculate CCFs for X Sgr. Line splitting manifests much more strongly in the case of weak metallic lines than in the case of strong. We confirm previous results that line splitting is not detected in Balmer lines.
\end{itemize}

\veloce\ time-series shed new light on the line splitting reported before in individual Cepheids. We were able to study the phenomenon with great detail. Contrary to expectations, line splitting is unlikely to originate as a result of pulsation-induced shock waves, but another mechanisms must be at play. Particularly interesting is the connection between the additional signals found in periodograms with the line splitting periodicity, and the possible link between the line splitting and average FWHM. The ongoing \veloce\ observations will allow to investigate these intriguing discoveries further.

\begin{acknowledgements}
This work was supported by the European Research Council (ERC) under the European Union’s Horizon 2020 research and innovation programme (Grant Agreement No. 947660). RIA is funded by the SNSF through an Eccellenza Professorial Fellowship, grant number PCEFP2\_194638.

This work uses frequency analysis software written by R. Smolec.

The Euler telescope is funded by the Swiss National Science Foundation (SNSF). Early \veloce\ observations ($2010-2016$) were enabled by SNSF project funding from grant Nos. 119778, 130295, and 140893.

This research is based on observations made with the Mercator Telescope, operated on the island of La Palma by the Flemish Community, at the Spanish Observatorio del Roque de los Muchachos of the Instituto de Astrof\'isica de Canarias. {\it Hermes} is supported by the Fund for Scientific Research of Flanders (FWO), Belgium, the Research Council of K.U. Leuven, Belgium, the Fonds National de la Recherche Scientifique (F.R.S.-FNRS), Belgium, the Royal Observatory of Belgium, the Observatoire de Gene\`eve, Switzerland, and the Th\"uringer Landessternwarte, Tautenburg, Germany.

We acknowledge the contributions of all observers who contributed to collecting the VELOCE dataset.
\end{acknowledgements}

\bibliographystyle{aa} 
\bibliography{bibliography.bib}

\end{document}